\pdfoutput=1
\documentclass[aps,prc,reprint,showpacs,floatfix,tightenlines,superscriptaddress]{revtex4-1}

\usepackage{amsmath}
\usepackage{txfonts}
\usepackage{graphicx}
\usepackage{color}

\usepackage[colorlinks=true,linkcolor=blue,citecolor=blue,urlcolor=blue]{hyperref}

\hypersetup{
  pdftitle={The Wigner function and short-range correlations in the deuteron},%
  pdfauthor={Thomas Neff <t.neff@gsi.de>}
}

\renewcommand{\vec}[1]{\boldsymbol{\mathbf{#1}}}
\newcommand{\unitvec}[1]{\hat{\boldsymbol{\mathbf{#1}}}}
\newcommand{\op}[1]{\widehat{#1}}
\newcommand{\conj}[1]{{#1}^*}

\newcommand{\bra}[1]{\bigl\langle {#1} \bigr\rvert}
\newcommand{\ket}[1]{\bigl\vert {#1} \bigr\rangle }
\newcommand{\expect}[1]{\bigl\langle {#1} \bigr\rangle} 
\newcommand{\braket}[2]{\bigl\langle {#1} \big\vert {#2} \bigr\rangle} 
\newcommand{\matrixe}[3]{\bigl\langle {#1} \big\vert\, {#2}\, \big\vert {#3} \bigr\rangle}

\newcommand{\cg}[6]{\left\langle \begin{matrix} {#1} & {#3} \\ {#2} & {#4} \end{matrix}\, 
\right\rvert\left. \begin{matrix} {#5} \\ {#6} \end{matrix} \right\rangle}
\newcommand{\sixj}[6]{\left\{ \begin{matrix} {#1} & {#2} & {#3} \\ {#4} & {#5} & {#6} \end{matrix} \right\} }
\newcommand{\ninej}[9]{\left\{ \begin{matrix} {#1} & {#2} & {#3} \\ {#4} & {#5} & {#6} \\ {#7} & {#8} & {#9} \end{matrix} \right\} }

\newcommand{\fm}{\ensuremath{\textrm{fm}}}

\newcommand{\nuc}[2]{$^{#1}${#2}}

\DeclareMathOperator{\Tr}{Tr}

\begin{document}

\title{The Wigner function and short-range correlations in the deuteron} 

\author{T. Neff}
\email{email: t.neff@gsi.de}
\affiliation{GSI Helmholtzzentrum f\"ur Schwerionenforschung GmbH,
Planckstra{\ss}e~1, 64291~Darmstadt, Germany}

\author{H. Feldmeier}
\affiliation{GSI Helmholtzzentrum f\"ur Schwerionenforschung GmbH,
Planckstra{\ss}e~1, 64291~Darmstadt, Germany}
\affiliation{Frankfurt Institute for Advanced Studies, Max-von-Laue
Stra{\ss}e~1, 60438~Frankfurt, Germany}

\date{\today}

\begin{abstract} 
  \begin{description}
   \item[Background] The deuteron shows the essential features of short-range correlations found in all nuclei. Experimental observables related to short-range correlations are connected with the high-momentum components of one- and two-body momentum distributions. An intuitive understanding of short-range correlations is provided by the suppression of the two-body density in coordinate space at small distances.
   \item[Purpose] The Wigner function provides a quasi-probability distribution in phase-space that allows to investigate short-range correlations as a function of distance and relative momentum in a unified picture. 
   \item[Method] The Wigner function for the deuteron is calculated for bare and SRG evolved AV8' and N3LO interactions and investigated as a function of distance, relative momentum and angular orientation. Partial momentum and coordinate space distributions are obtained by integrating over parts of phase space.	
   \item[Results] The Wigner function shows a pronounced low-momentum peak that is not affected by short-range correlations and a high-momentum shoulder at small distances that reflects short-range correlations. Oscillations of the Wigner function are related to interference of low- and high-momentum components.       
   \item[Conclusions] Short-range correlations are a truly quantum-mechanical phenomenon caused by interference of low- and high-momentum components in the wave function.
  \end{description} 
\end{abstract}


\maketitle

\section{Introduction}
\label{sec:intro}

The quantum analogue to the classical phase space distribution introduced by Wigner \cite{wigner32} has been investigated in many areas for a better semi-classical perception of quantal systems. Reviews of Wigner and related phase space distributions are given in \cite{hillery84,lee95,case08}.

Wigner functions are quite popular in particle physics as a tool to understand nucleon parton distributions \cite{xi04,belitsky04,lorce12}. In the context of nuclear structure Wigner functions have been studied for shell model wave functions \cite{shlomo81,prakash81} but to our knowledge only for effective and not for realistic interactions. On the other hand the issue of short-range correlations induced by realistic interactions has been studied widely. Short-range correlations can be investigated in coordinate space where they are reflected in a suppression of the two-body density at short distances or in momentum space where they manifest themselves in high-momentum components of one- and two-body densities. This short-range physics is usually contrasted with the low momentum part that contains mainly long-range correlations which are describable by superpositions of a smaller number of uncorrelated Slater determinants.

Different from considering of the coordinate density and momentum density separately, the Wigner distribution provides explicitly relations between the position and momentum. Therefore we believe that the phase space nature  of the Wigner function furnishes a better intuitive picture and helps in understanding how long- and short-range or low- and high-momentum physics comes together. On the other hand it has been shown in many publications \cite{forest96,schiavilla07,wiringa14,alvioli12,ciofi15,src11,src15b} that the short range correlations which are induced by the nucleon-nucleon interaction show a universal behavior and look very similar in all nuclei including the deuteron. To keep everything simple we therefore restrict ourselves in this paper to the deuteron where the full information about the system is contained in the wave function for the relative motion and explore in how far the Wigner distribution provides additional insight into the contributions of short- and long-range physics.

Nucleons are the effective degrees of freedom for low energy nuclear physics and they interact through an effective nucleon-nucleon force. The long range part of the nucleon-nucleon interaction is determined by pion exchange. At intermediate distances two pion exchange becomes dominant. The short distance behavior is modeled in different ways. The Argonne $v_{18}$ (AV18) and $v_8'$ (AV8') interactions \cite{wiringa95} use a local phenomenological form, while interactions derived from chiral perturbation theory employ regulators. The popular Idaho interaction \cite{entem03} derived from chiral perturbation theory (N3LO) uses non-local cut offs in momentum space. Recently the choice and form of regulators has been discussed intensively \cite{epelbaum15,epelbaum15b}. In particular inconsiderate manipulations of momentum space matrix elements can lead to highly non-local interactions as has been shown in Ref.~\cite{feldmeier14} where a phase-space representation of the nucleon-nucleon interaction has been presented.

Independent of the choice of regulators all these interactions describe the nucleon-nucleon scattering data up to pion production threshold and the deuteron properties equally well (their parameters are fitted to the scattering data). However due to the different choices of regulators these interactions show a quite different behavior at short distances or high momenta. These differences are reflected in the short-range correlations obtained with different interactions. In Ref.~\cite{src15b} two-body densities in coordinate and momentum space in \nuc{4}{He} for the AV8' and chiral N3LO interaction have been investigated. 

In nuclear many-body calculations a special treatment of short-range correlations is necessary to make a solution of the many-body problem possible. This can be achieved by either including short-range correlations explicitly, for example with Jastrow correlation functions, or by employing renormalized interactions. The similarity renormalization group (SRG) is a very popular approach that allows to soften the interaction by means of a unitary transformation \cite{bogner07,bogner10,ucom10}. By using SRG transformed interactions the high-momentum components in the wave function get eliminated. However the short-range or high-momentum information is not lost and can be recovered by correspondingly transforming observables (density operators). This has been shown in the deuteron \cite{anderson10,more15} and in \nuc{4}{He} \cite{src15b}.

In this paper we will investigate the Wigner function of the deuteron for the AV8' and N3LO interactions. Comparing with the Wigner function obtained with the SRG softened interactions allows to isolate the short-range correlation effects. In Sec.~\ref{sec:method} we will discuss the Wigner function and how it can be calculated for the deuteron. In Sec.~\ref{sec:results} we discuss the Wigner function, reduced Wigner functions and partial coordinate space and momentum space distributions obtained by integrating the Wigner function over ranges of distances or relative momenta. Furthermore we discuss tensor correlations in coordinate and momentum space. We give a summary and conclusions in Sec.~\ref{sec:summary}. Technical details about the calculation of the Wigner function are given in the Appendix.

\section{Method}
\label{sec:method}

\subsection{Wigner function}

The Wigner function introduced in 1932 by Wigner aimed at finding a quantum phase-space distribution that depends on both, coordinate and momentum and becomes the classical phase-space distribution in the limit of large particle numbers or where detailed phases in the many-body system do not matter anymore. For a system with a single degree of freedom described by a wave function it is well known that the Heisenberg uncertainty relation puts limits on the classical interpretation of the Wigner function, like the appearance of areas where it is negative. 

Let $\op{\rho}$ be a density that describes a single degree-of-freedom (e.g. the one-body density). Then the expectation value of an one-body observable $\op{O}$ is given by
\begin{align}
  \expect{\op{O}} 
  &= \Tr \big\{ \op{\rho} \, \op{O} \big \} \\
  &= \int d^3r' d^3r'' \: \matrixe{\vec{r}'}{\op{\rho}}{\vec{r}''} \matrixe{\vec{r}''}{\op{O}}{\vec{r}'} \\
 	&= \int d^3r \: d^3s \: \matrixe{\vec{r}+\tfrac{1}{2}\vec{s}}{\op{\rho}}{\vec{r}-\tfrac{1}{2}\vec{s}}
		\matrixe{\vec{r}-\tfrac{1}{2}\vec{s}}{\op{O}}{\vec{r}+\tfrac{1}{2}\vec{s}} \label{eq:w2}\\
	&= \int d^3r \: d^3p \: \frac{1}{(2\pi)^3} \int d^3s \: \matrixe{\vec{r}+\tfrac{1}{2}\vec{s}}{\op{\rho}\,}{\vec{r}-\tfrac{1}{2}\vec{s}} \: e^{-i \vec{p}\cdot\vec{s}} \nonumber\\
 	&\quad\quad \times 
 \int d^3\!s' \: \matrixe{\vec{r}-\tfrac{1}{2}\vec{s}'}{\op{O}}{\vec{r}+\tfrac{1}{2}\vec{s}'} \: e^{i \vec{p}\cdot\vec{s}'}\label{eq:w3} \: .       
\end{align}
One arrives at Eq.~\eqref{eq:w2} by taking the trace in coordinate representation and changing variables to $\vec{r}=(\vec{r}'+\vec{r}'')/2$ and $\vec{s}=\vec{r}'-\vec{r}''$. The insertion of the $\delta$-function
\begin{equation}
	\delta^3(\vec{s}'-\vec{s}) = \frac{1}{(2\pi)^3}\int d^3p \: e^{i\vec{p}\cdot(\vec{s}'-\vec{s})}\end{equation}
leads to Eq.~\eqref{eq:w3}. Introducing the Wigner representation (or Wigner function) $W(\vec{r},\vec{p})$ of the density operator $\op{\rho}$
\begin{align}\label{eq:wigner0-r}
	W(\vec{r},\vec{p}) &= \frac{1}{(2\pi)^3}\int d^3s \: \matrixe{\vec{r}+\tfrac{1}{2}\vec{s}}{\op{\rho}}{\vec{r}-\tfrac{1}{2}\vec{s}}\:e^{-i \vec{p}\cdot\vec{s}}\\
 &= \frac{1}{(2\pi)^3}\int d^3 \varkappa \: \matrixe{\vec{p}+\tfrac{1}{2}\vec{\varkappa}}{\op{\rho}\,}{\vec{p}-\tfrac{1}{2}\vec{\varkappa}}\:e^{i \vec{\varkappa}\cdot\vec{r}} \label{eq:wigner0-p}
\end{align}
and the Weyl representation $O_W(\vec{r},\vec{p})$ \cite{weyl31} of the operator $\op{O}$
\begin{align} \label{eq:weyl-r}
	O_W(\vec{r},\vec{p}) &= \int d^3s' \:  
		\matrixe{\vec{r}-\tfrac{1}{2}\vec{s}'}{\op{O}}{\vec{r}+\tfrac{1}{2}\vec{s}'}\: e^{i \vec{p}\cdot\vec{s}'}\\
             &= \int d^3\varkappa \: \matrixe{\vec{p}-\tfrac{1}{2}\vec{\varkappa}}{\op{O}}{\vec{p}+\tfrac{1}{2}\vec{\varkappa}}\:e^{-i \vec{\varkappa}\cdot\vec{r}} \label{eq:weyl-p}
\end{align}
one arrives at the expression
\begin{equation}\label{eq:expect}
	\expect{\op{O}} = \int d^3r \: d^3p \: W(\vec{r},\vec{p}) \: O_W(\vec{r},\vec{p})                  
\end{equation}
which looks like the classical  expression when identifying the Wigner function $W(\vec{r},\vec{p})$ as the phase space distribution and the Weyl representation $O_W(\vec{r},\vec{p})$ as the classical observable depending on $\vec{r}$ and $\vec{p}$. But Eq.~\eqref{eq:expect} is still quantum mechanically exact. Eqs.~\eqref{eq:wigner0-p} and \eqref{eq:weyl-p} give the Wigner and  Weyl representation in terms of the momentum basis, respectively.

For any local operator, i.e. $\op{O}=O(\op{\vec{r}})$ the matrix elements in Eq.~\eqref{eq:weyl-r} are diagonal and hence the Weyl representation is just the classical analogue $O_W(\vec{r},\vec{p}) = O(\vec{r})$, not depending on $\vec{p}$. With Eq.~\eqref{eq:wigner0-p} one sees that the analogue holds true for an observable $\op{O} = O(\op{\vec{p}})$ that depends only on the momentum operator $\op{\vec{p}}$, here $O_W(\vec{r},\vec{p}) = O(\vec{p})$. For products of noncommuting operators like $\op{x}$ and $\op{p}_x$ the correspondence need not be one-to-one. In particular the reverse mapping going from a classical observable $O_W(\vec{r},\vec{p})$ to the corresponding operator $O(\op{\vec{r}},\op{\vec{p}})$ is subtle \cite{hillery84}.

\subsection{Deuteron wave function}

The full information about the deuteron is contained in its eigenstate $\ket{\Psi;JM}$. Due to the properties of the nuclear force the state with $J=0$ is not bound and the only bound state is found in the $J=1$ channel. In this channel we have total spin $S=1$ and strong contributions from the tensor force which couples orbital angular momentum $L=0$ and $L=2$ with the total spin $S=1$ to total angular momentum $J=1$. The wave function of the deuteron is thus given by
\begin{multline}
	\braket{\vec{r};SM_S}{\Psi;JM} = \\ 
	\sum_{L M_L} \cg{L}{M_L}{S}{M_S}{J}{M} \psi^{JL}(r) Y_{LM_L}(\unitvec{r}) \: .
\end{multline}

In this paper we expand the wave functions using Gaussian basis functions 
\begin{multline}
	\braket{\vec{r};SM_S}{\Psi;JM} = \sum_{L M_L} \cg{L}{M_L}{S}{M_S}{J}{M} \\
	\times \left( \sum_k r^L \exp\left\{-\frac{\vec{r}^2}{2 a_k}\right\} \psi^{JL}_k \right) Y_{LM_L}(\unitvec{r})
	\label{eq:gaussian}
\end{multline}
where $a_k$ are the width parameters. The Gaussian basis functions have been chosen because they make the calculation of the Wigner function simpler as will be explained later.

We solve the eigenvalue problem using the Gaussian basis $a_k = a_0 \times 2^k$ with $a_0 = 0.01\:\fm^2$ and $k=0,1,\ldots,13$. 

\begin{figure}
	\includegraphics[width=\columnwidth]{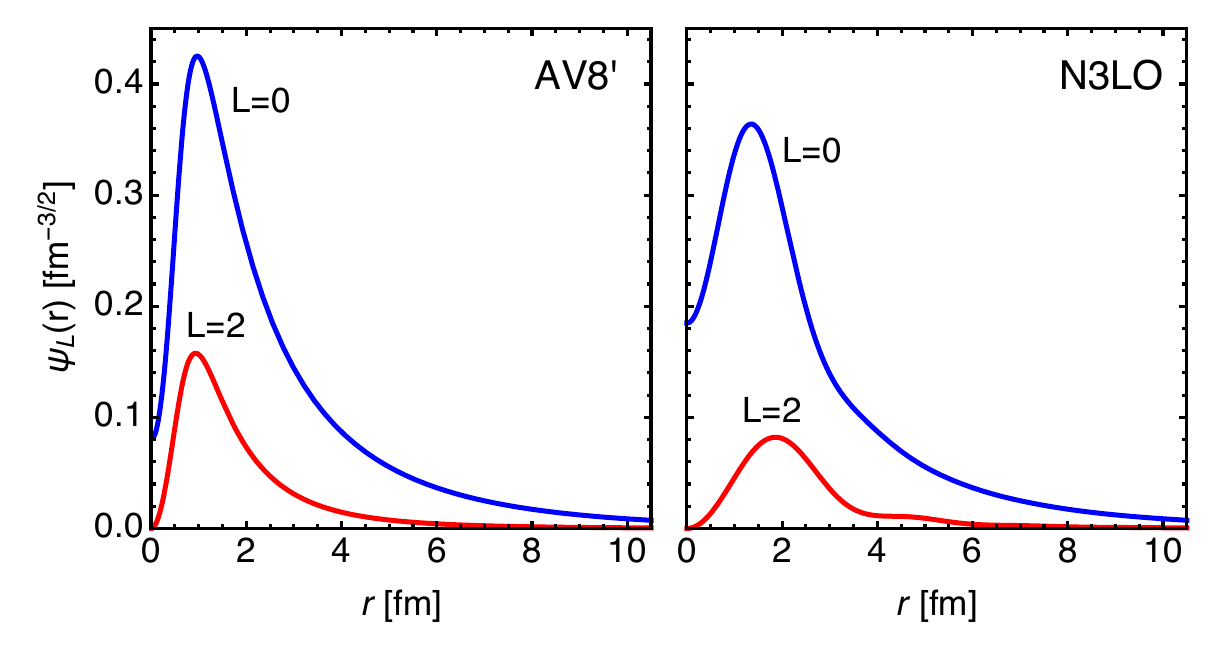}\\
	\includegraphics[width=\columnwidth]{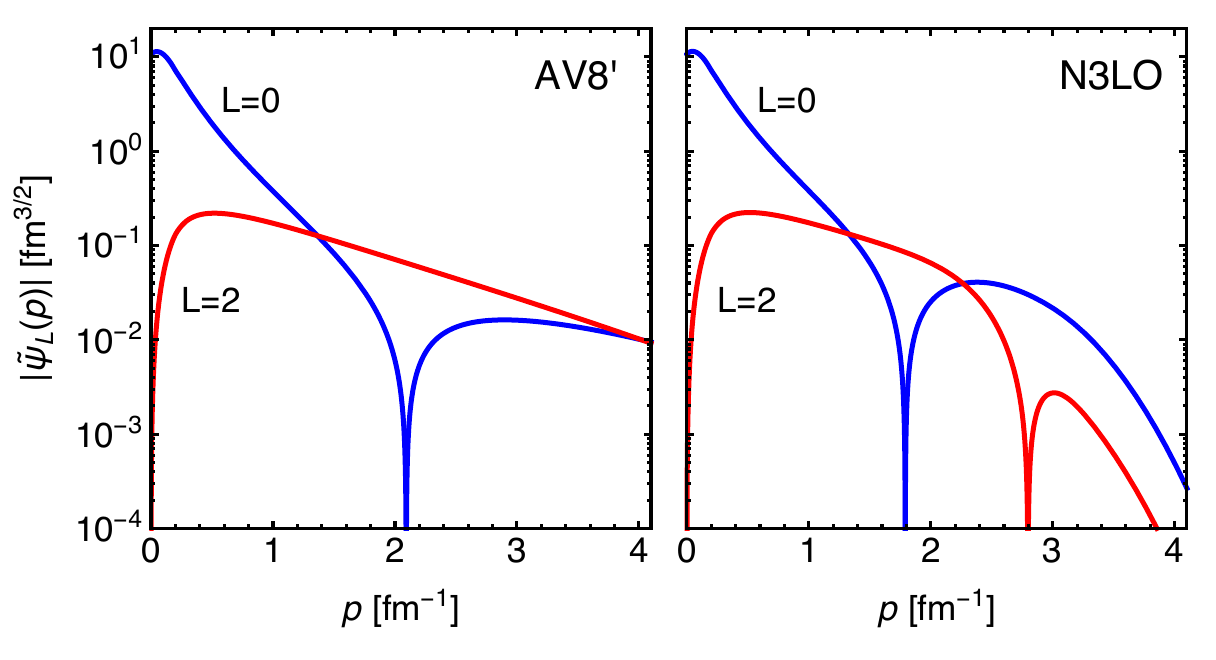}
	\caption{(Color online) $L=0$ and $L=2$ components of the deuteron wave functions in coordinate (top) and momentum space (bottom) for the AV8' (left) and N3LO (right) interactions.}
	\label{fig:wavefunctions}
\end{figure}

The deuteron wave functions in coordinate and momentum space for AV8' and N3LO interactions are shown in Fig.~\ref{fig:wavefunctions}. The stronger short-range repulsion of the AV8' interaction is reflected in the more pronounced suppression of the coordinate space wave function at small distances and in the tails of the momentum space wave function. The AV8' interaction also has in addition a `harder' tensor force, the $L=2$ high momentum components extend to higher momenta compared to those obtained with the N3LO interaction. The N3LO wave functions also reflect the properties of the regulator. The kink in the coordinate space wave function at distances of about 4~fm shows that the long-range behavior of the interaction is changed by the regulator. A different regularization scheme for the chiral interactions with local regulators for the pion exchanges does not show such a behavior \cite{epelbaum15b}. 

\subsection{Wigner function}

One can also express the information about the deuteron in the density matrix for a pure state with definite orientation
\begin{equation}
  \op{\rho}_M = \ket{\Psi;JM}\bra{\Psi;JM} \: .
\end{equation}
In the case that we have no information about the orientation of the deuteron (unpolarized) the density matrix is
\begin{equation}
  \op{\rho} = \frac{1}{2J+1} \sum_M \ket{\Psi;JM}\bra{\Psi;JM} \: .
  \label{eq:densitymatrix}
\end{equation} 
In this case the density operator is a scalar operator and all observables will be invariant under rotations. In the following we will always discuss this case.

The most general Wigner function that contains all information is given by a matrix in spin space
\begin{multline}
	W_{M_S,{M_S}'}(\vec{r},\vec{p}) = \\
		\frac{1}{(2\pi)^3} \int d^3s \: 
		\matrixe{\vec{r}+\tfrac{1}{2}\vec{s};SM_S}{\op{\rho}}{\vec{r}-\tfrac{1}{2}\vec{s};S{M_S}'} 
		\: e^{-i \vec{p}\cdot\vec{s}} \: .
	\label{eq:wigner}
\end{multline}

The calculation of the Wigner function is not trivial as the integration over $\vec{s}$ deals with three independent vectors $\vec{r}$, $\vec{p}$ and $\vec{s}$. Expanding the wave function with Gaussian basis functions simplifies the integration over $\vec{s}$. For a pure $S$-wave function it becomes a simple Gaussian integration. In Appendix~\ref{app:calculation} we give some more details.  

The general matrix valued Wigner function can be reduced by taking the trace in spin space
\begin{equation}
	W(\vec{r},\vec{p}) = \sum_{M_S} W_{M_S,M_S}(\vec{r},\vec{p})
\end{equation}
where the sum is over all spin orientations. Thus it contains no information about the spin orientation any more. As there is no preferred orientation, $W(\vec{r},\vec{p})$ only depends on $r=|\vec{r}|$, $p=|\vec{p}|$ and the angle between $\vec{r}$ and $\vec{p}$, $\cos \vartheta = (\vec{r}\cdot\vec{p})/(r p)$. If we are not interested in the angular correlations between $\vec{r}$ and $\vec{p}$ we can further reduce the Wigner function to 
\begin{equation}
	\begin{split}
		W(r,p) & = \int d\Omega_r \int d\Omega_p W(\vec{r},\vec{p}) \\
		& = 8 \pi^2 \int d(\cos \vartheta) W(r,p,\cos \vartheta) \: .
	\end{split}
\end{equation}

As mentioned earlier the most general Wigner function matrix contains the full information about the system and we can calculate for example coordinate space densities
\begin{equation}
	\rho_{M_S}(\vec{r}) = \matrixe{\vec{r};SM_S}{\op{\rho}}{\vec{r};SM_S} = \int d^3p \: W_{M_S,M_S}(\vec{r},\vec{p}) 
\end{equation}
and momentum distributions
\begin{equation}
	n_{M_S}(\vec{p}) = \matrixe{\vec{p};SM_S}{\op{\rho}}{\vec{p};SM_S} = \int d^3r \: W_{M_S,M_S}(\vec{r},\vec{p})
\end{equation}
from the Wigner function.

Without information about the spin orientation the coordinate space densities and momentum distributions are scalar objects:
\begin{equation}
	\rho(r) = \int d\Omega_r \sum_{M_S} \rho_{M_S}(\vec{r}) = \int dp \: p^2 W(r,p)
\end{equation}
\begin{equation}
	n(p) = \int d\Omega_p \sum_{M_S} n_{M_S}(\vec{p}) = \int dr \: r^2 W(r,p)
\end{equation}

The Wigner function can of course also be used to calculate the full off-diagonal density matrix
\begin{equation}
	\begin{split}
  \rho_{M_S,{M_S}'}(\vec{r};\vec{r}') & = 
  \matrixe{\vec{r};SM_S}{\op{\rho}}{\vec{r}';SM_S'} \\
  & = \int d^3p \: W_{M_S,{M_S}'}(\tfrac{1}{2}(\vec{r}+\vec{r}'), \vec{p}) \: e^{i \vec{p} \cdot (\vec{r}-\vec{r}')} \: .
  \end{split}
\end{equation}

\section{Results}
\label{sec:results}

\subsection{Reduced Wigner function $W(r,p)$}

\begin{figure}
	\centering
  \includegraphics[width=0.33\textwidth]{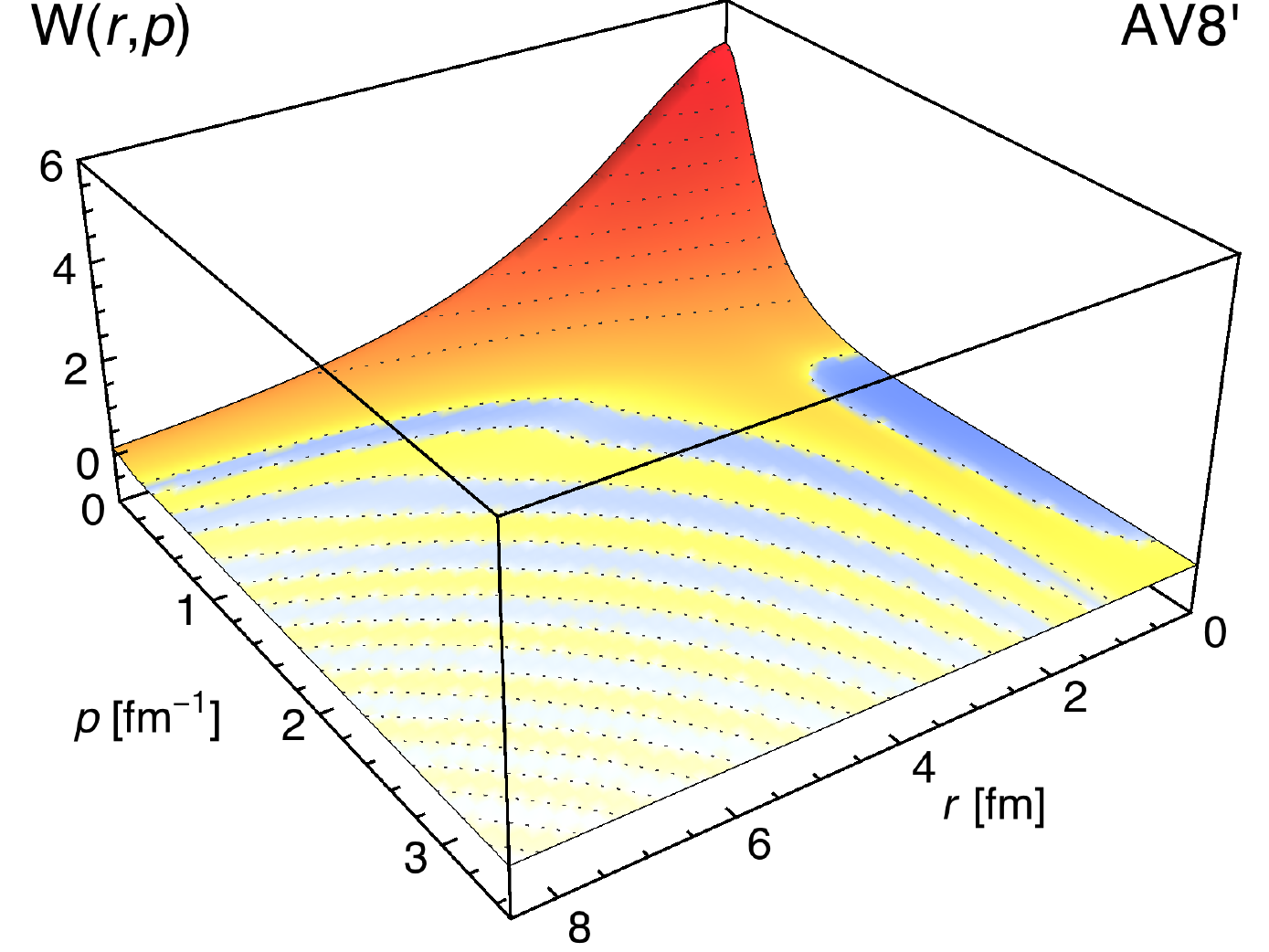}\\
  \includegraphics[width=0.33\textwidth]{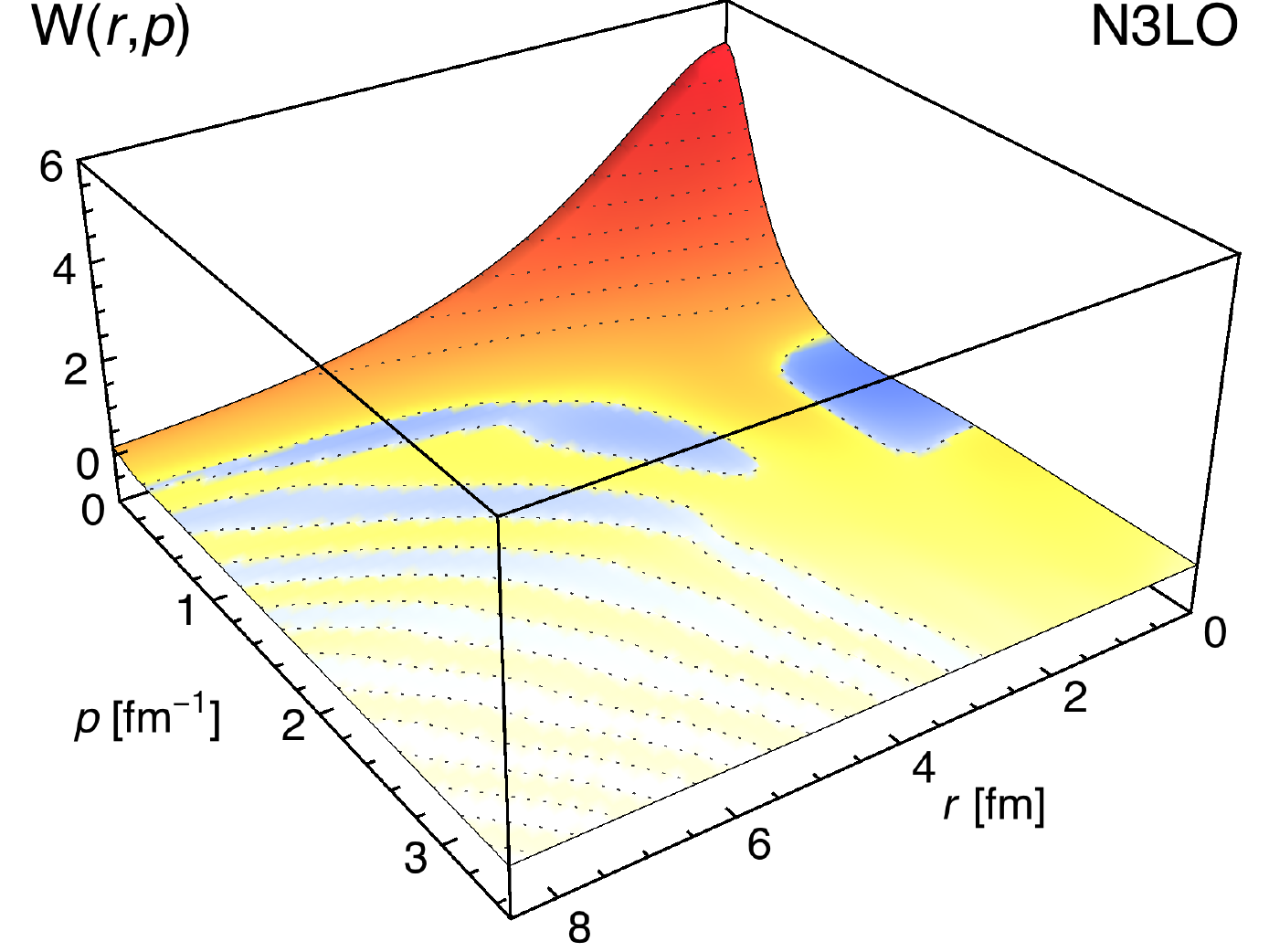}
	\caption{(Color online) Wigner function $W(r,p)$ of the deuteron for the AV8' (top) and N3LO (bottom) interactions. Yellow and red colors indicate positive values, blue colors negative values of the Wigner function.}
	\label{fig:wignerrp}
\end{figure}

\begin{figure}
	\centering
	\includegraphics[width=0.33\textwidth]{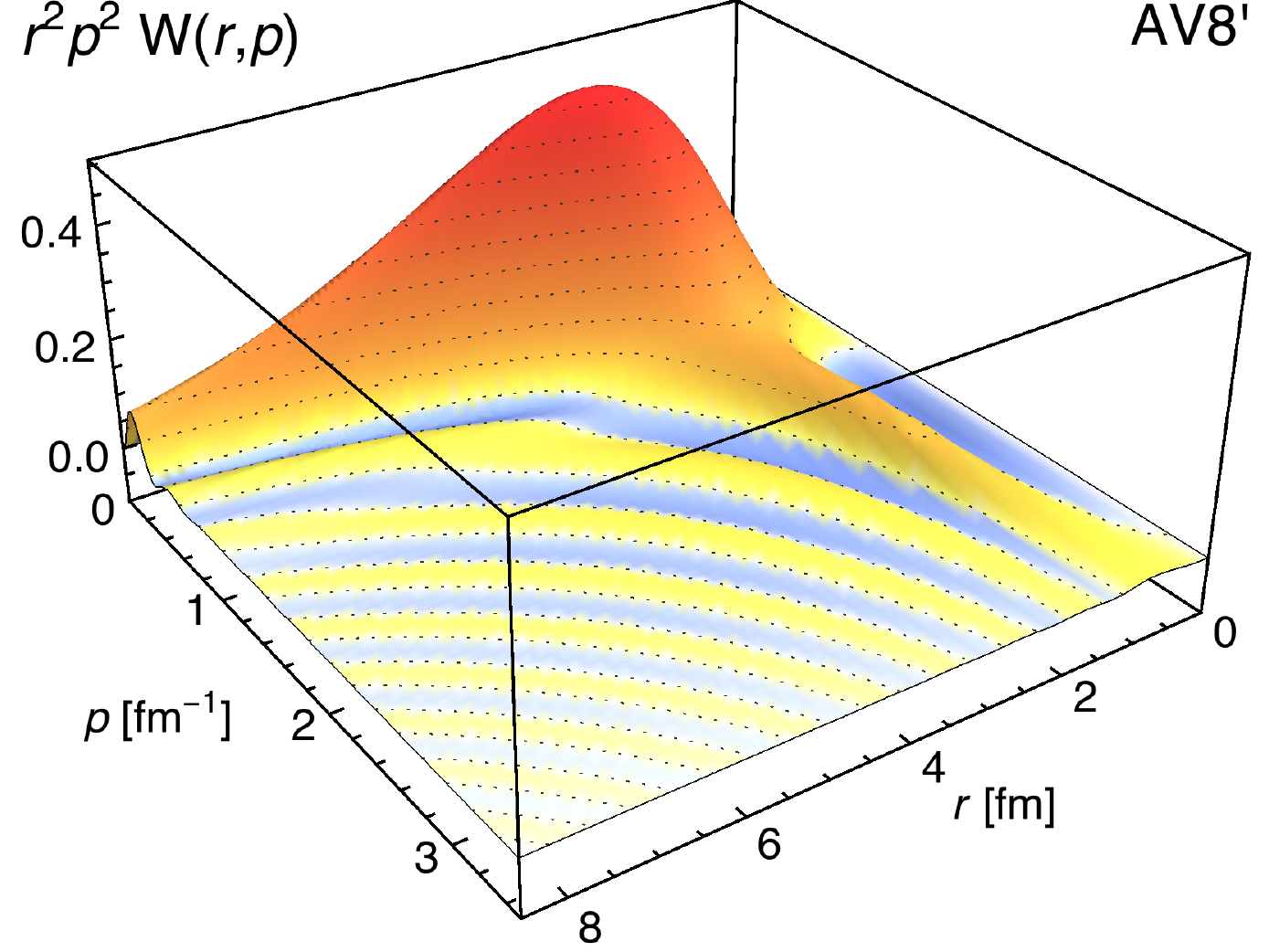}\\    
	\includegraphics[width=0.33\textwidth]{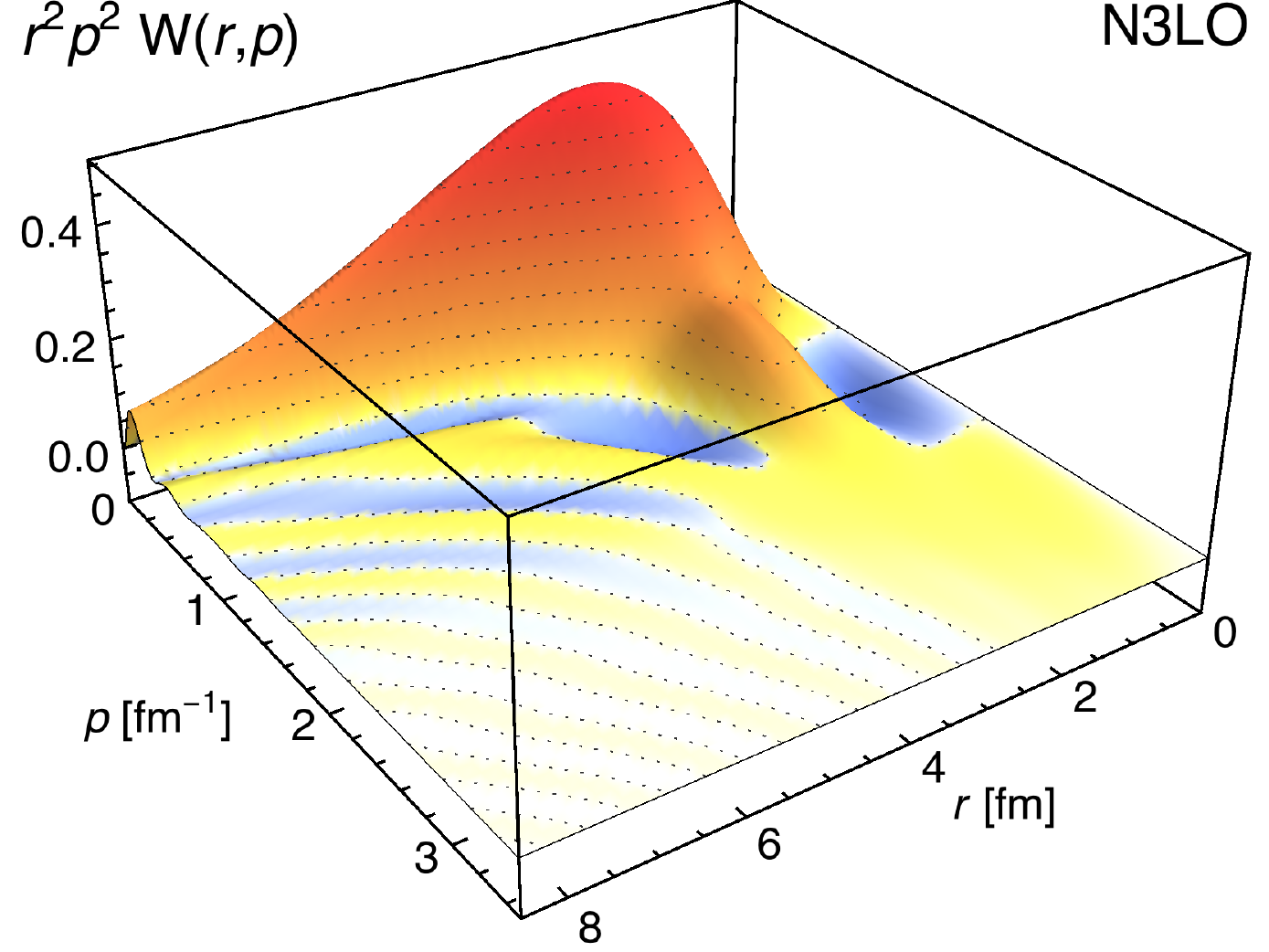}
	\caption{(Color online) Wigner function scaled with phase space volume element $r^2 p^2 W(r,p)$ of the deuteron for the AV8' (top) and N3LO (bottom) interactions.}
	\label{fig:wignerrpscaled}
\end{figure}

The reduced Wigner function $W(r,p)$ only depends on the absolute values of distance and relative momentum and hides or averages out spin information that is contained in the full Wigner function matrix $W_{M_S,{M_S}'}(\vec{r},\vec{p})$.

The Wigner functions obtained with the AV8' and N3LO interactions are shown in Fig.~\ref{fig:wignerrp}. The Wigner function contains the combined information about coordinate and momentum space and can be considered as the analogue of the classical phase space distribution. Contrary to a classical phase space distribution the Wigner function can become negative which makes clear that it can not be interpreted as a true probability distribution. The non-positivity reflects Heisenberg's uncertainty principle which does not allow to determine position and momentum with arbitrary precision at the same time.
 
In the following we will nevertheless treat the Wigner function as a kind of probability distribution and argue about ``contributions from small or large distances to the momentum distribution'' or ``low and high momentum components of the coordinate space density''. This is not meant in the sense of an observable quantity. Observable probabilities are always traces and integrals over large enough (given by the uncertainty relation) phase space volumes. The discussions presented in the following will however disentangle how specific phase space regions contribute to the total result.

The Wigner function shows some characteristic and maybe surprising features. The first observation is that the Wigner function takes its maximal values at $r=0$ although the coordinate space wave function is suppressed due to the repulsive core at small distances. Apart from the short distance region the Wigner function $W(r,p=0)$ falls off slowly as one would expect from the deuteron wave function. If we follow the Wigner function along the momentum axis $W(r=0,p)$ it falls off rapidly and becomes negative at $p \gtrsim 1\:\fm^{-1}$. This can be interpreted as the effect of the repulsive core. The suppression of the density at small distances is therefore caused by a cancellation of the positive contributions from small momenta with those from large momenta. If we look at larger distances of about $1 - 1.5\:\fm$ the Wigner function stays positive even for large relative momenta. 

The probability densities in coordinate and momentum space are linked to the Wigner function by
\begin{equation}
	\rho(r) = \int dp \: p^2 W(r,p), \quad 
	n(p) = \int dr \: r^2 W(r,p) \: .
\end{equation}
To relate the Wigner function to these `projections' on the coordinate or momentum axes it is therefore advantageous to show the Wigner function scaled with the phase space volume element $r^2 p^2$. This is displayed in Fig.~\ref{fig:wignerrpscaled}. These scaled Wigner functions can be interpreted as the quasi-probability distribution to find the nucleons at distance $r$ and relative momentum $p$. In this view the Wigner function is dominated by a peak that extends to large distances but only small momenta. This can be identified as the low-momentum contribution. Apart from this dominating peak there is a shoulder extending to large relative momenta for distances of about $1 - 1.5\:\fm$. This shoulder reflects the correlated part of the wave function that has its origin in the short-range correlations.

At larger distances and momenta the Wigner function shows oscillations that roughly follow a $r \cdot p \approx \mathrm{const}$ pattern. These oscillations have their origin in the interference of the short-range and the long-range components of the wave function. This is exemplified and explained in Sec.~\ref{app:schematic} where the Wigner function of a schematic wave function given by a superposition of a short- and long-range Gaussian is investigated.

\subsection{Wigner function for SRG evolved interaction}

\begin{figure}
	\centering
	\includegraphics[width=0.33\textwidth]{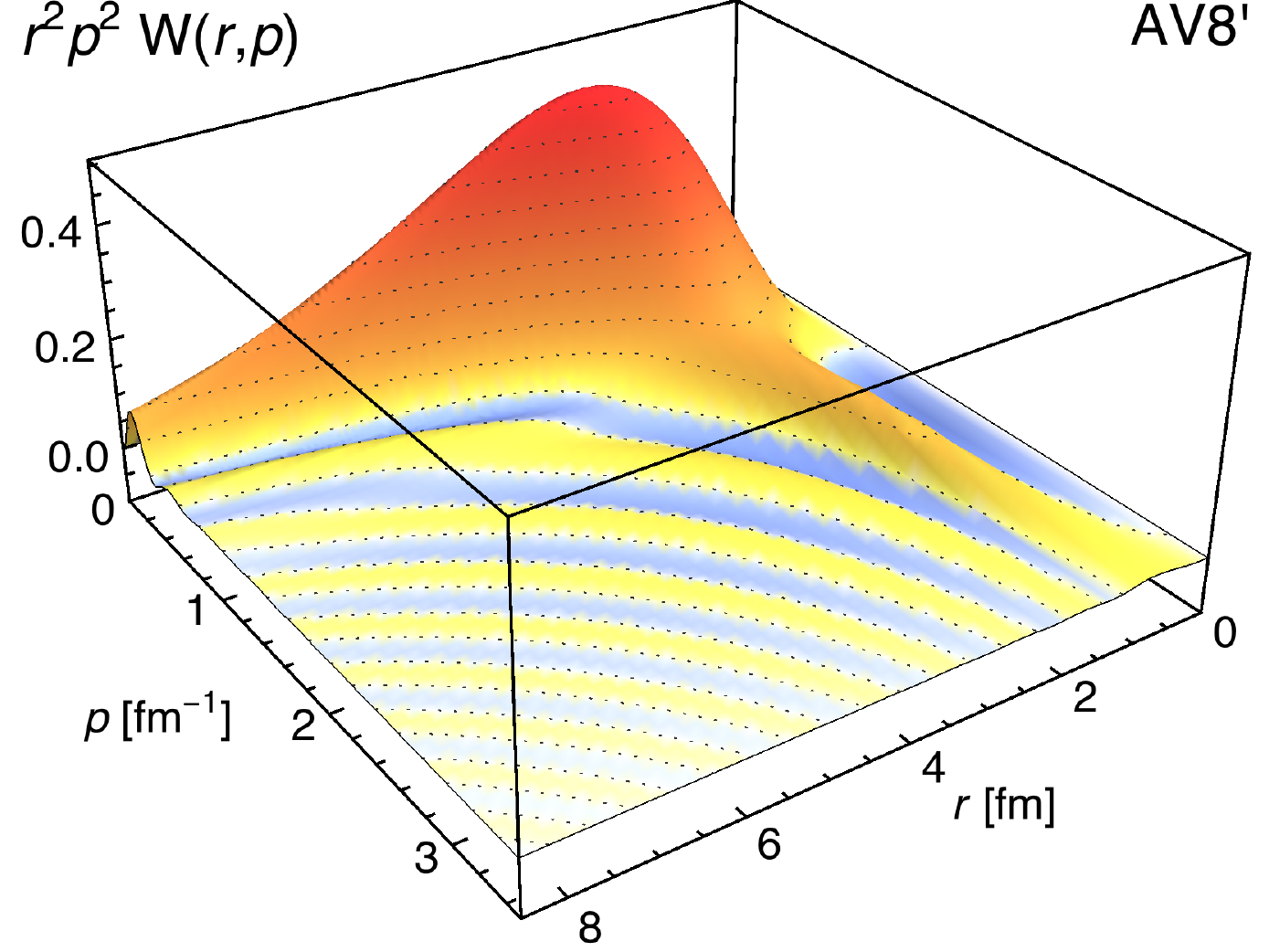}\\
	\includegraphics[width=0.33\textwidth]{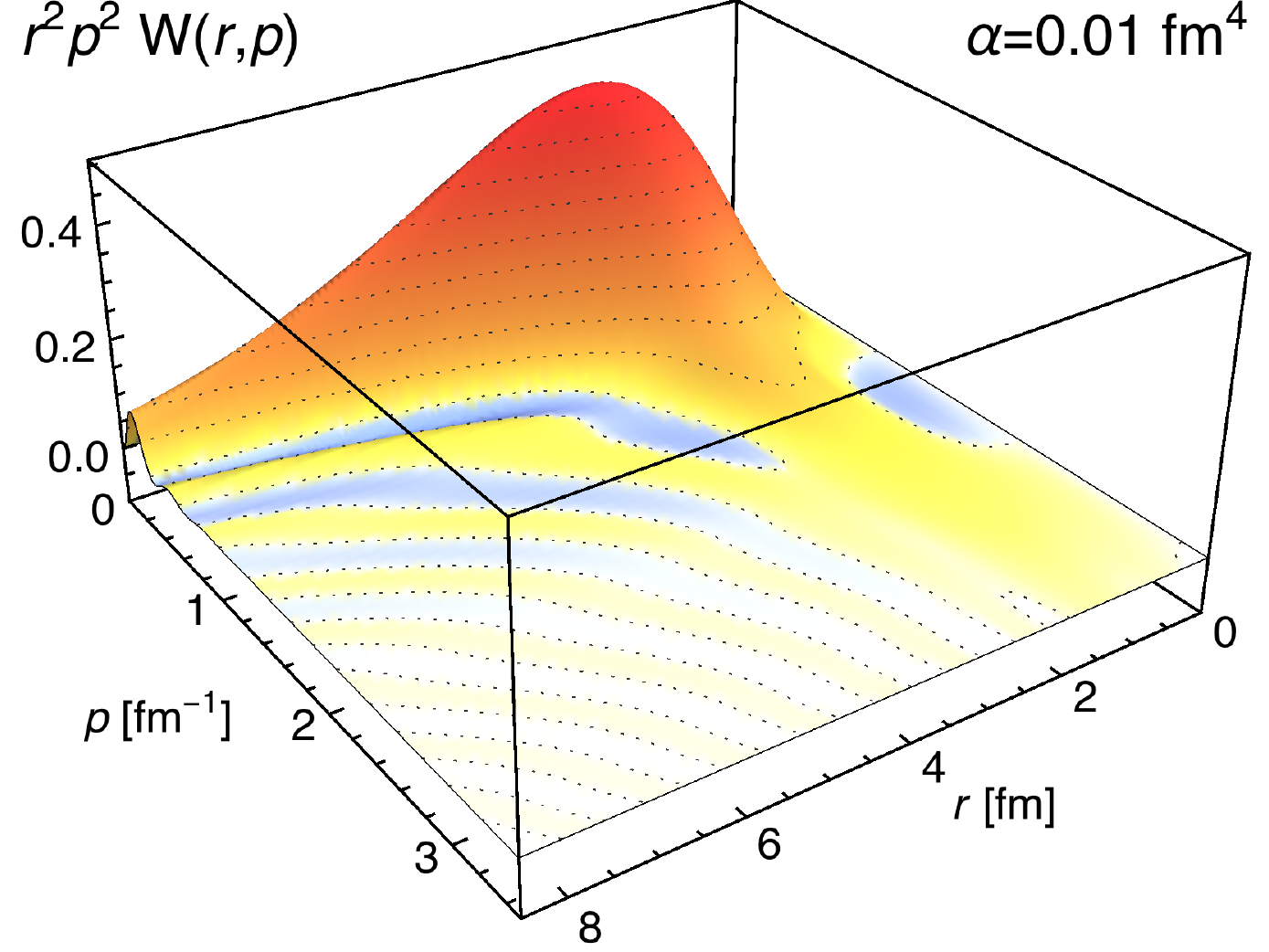}\\
	\includegraphics[width=0.33\textwidth]{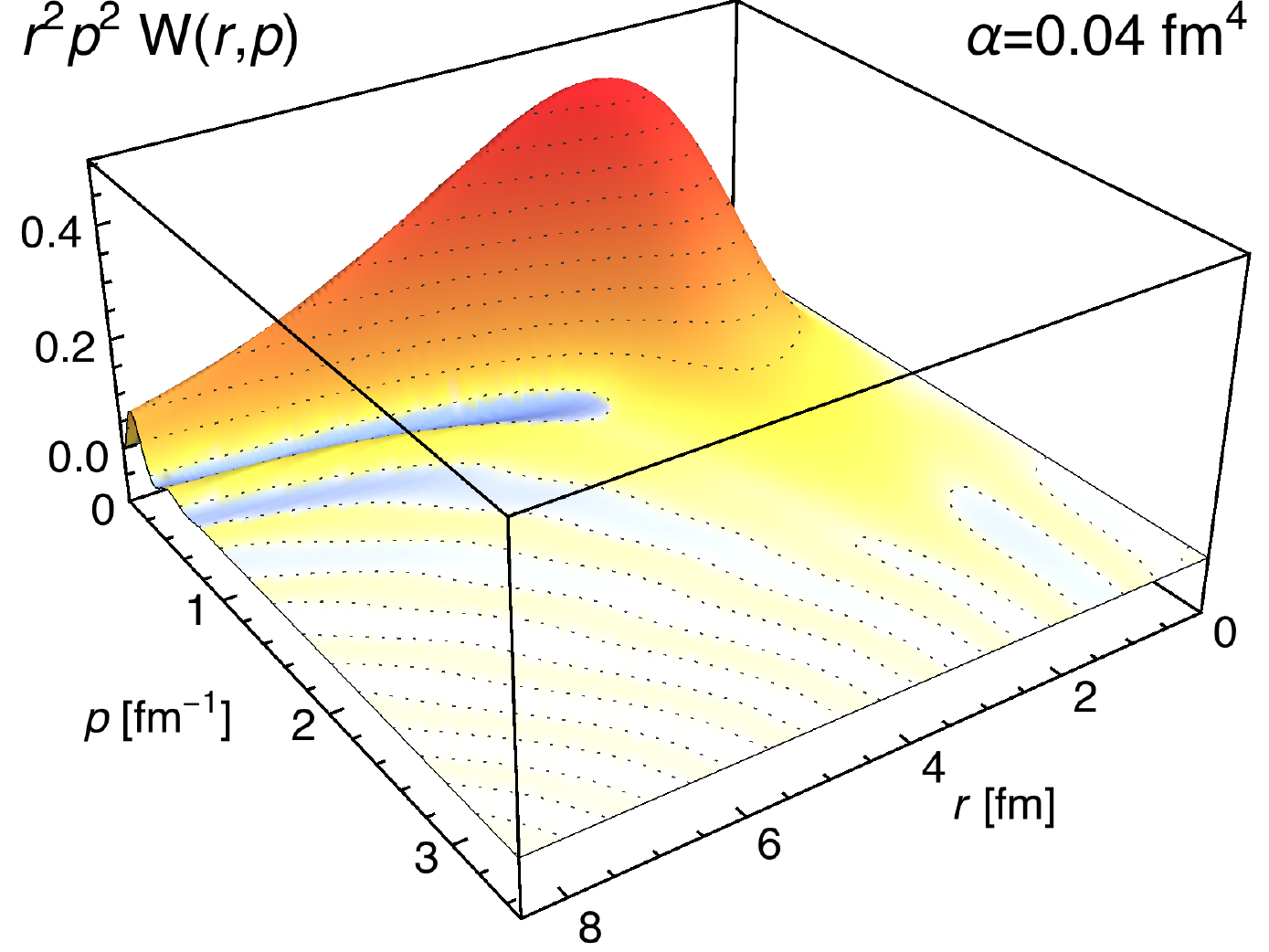}\\
	\includegraphics[width=0.33\textwidth]{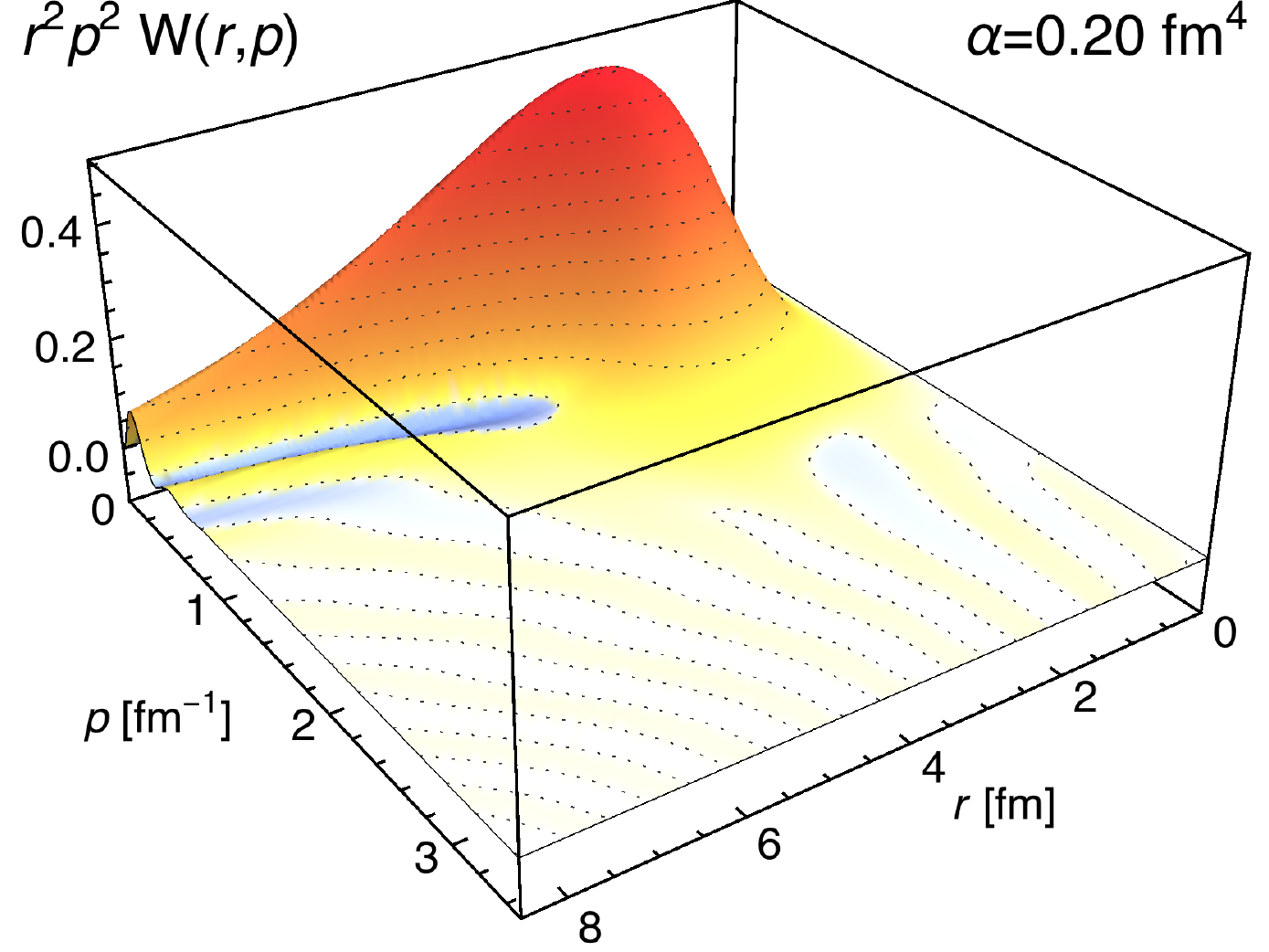}
	\caption{(Color online) Wigner function of the deuteron scaled with phase space volume element $r^2 p^2 W(r,p)$ for the SRG evolved AV8' interaction (bare, $\alpha {=} 0.01\:\fm^4$, $\alpha {=} 0.04\:\fm^4$, $\alpha {=} 0.20\:\fm^4$). }
	\label{fig:wignerscaledrp-av8p-srg}
\end{figure}

The similarity renormalization group (SRG) is used to transform realistic interactions into soft realistic interactions \cite{bogner07,bogner10,ucom10}. With increasing flow parameter $\alpha$ the interaction is more and more softened. The effect of the SRG evolution on the two-body densities in coordinate and momentum space has already been investigated in \cite{src15b}. With increasing flow parameter $\alpha$ the suppression of the two-body density at short-distances and the high momentum components in the two-body momentum distribution get eliminated. In Fig.~\ref{fig:wignerscaledrp-av8p-srg} we show the Wigner function for the SRG evolved AV8' interaction. It can be seen that the low-momentum part of the Wigner function ($p \lesssim 1.0\:\fm^{-1}$) is essentially unchanged not only at large but also at small distances as a function of the flow parameter. On the other hand the high-momentum region is strongly affected by the SRG transformation. The high momentum shoulder that can be seen in the Wigner function of the bare interaction gets more and more washed out with increasing flow parameter $\alpha$. We can also observe a change in the characteristics of the mid-momentum region ($1\:\fm^{-1} \lesssim p \lesssim 2\:\fm^{-1}$). For the bare interaction the Wigner function shows a strongly oscillating behavior that is more and more damped with increasing flow parameter $\alpha$, indicating that the interference between short- and long-range parts weakens because of the decreasing strength of the short-range part as can been easily seen for the schematic wave function discussed in Sec.~\ref{app:schematic}.

\subsection{Partial momentum distributions $n_<(p)$ and $n_>(p)$}
\label{sec:momentumsep}

\begin{figure}
	\includegraphics[width=\columnwidth]{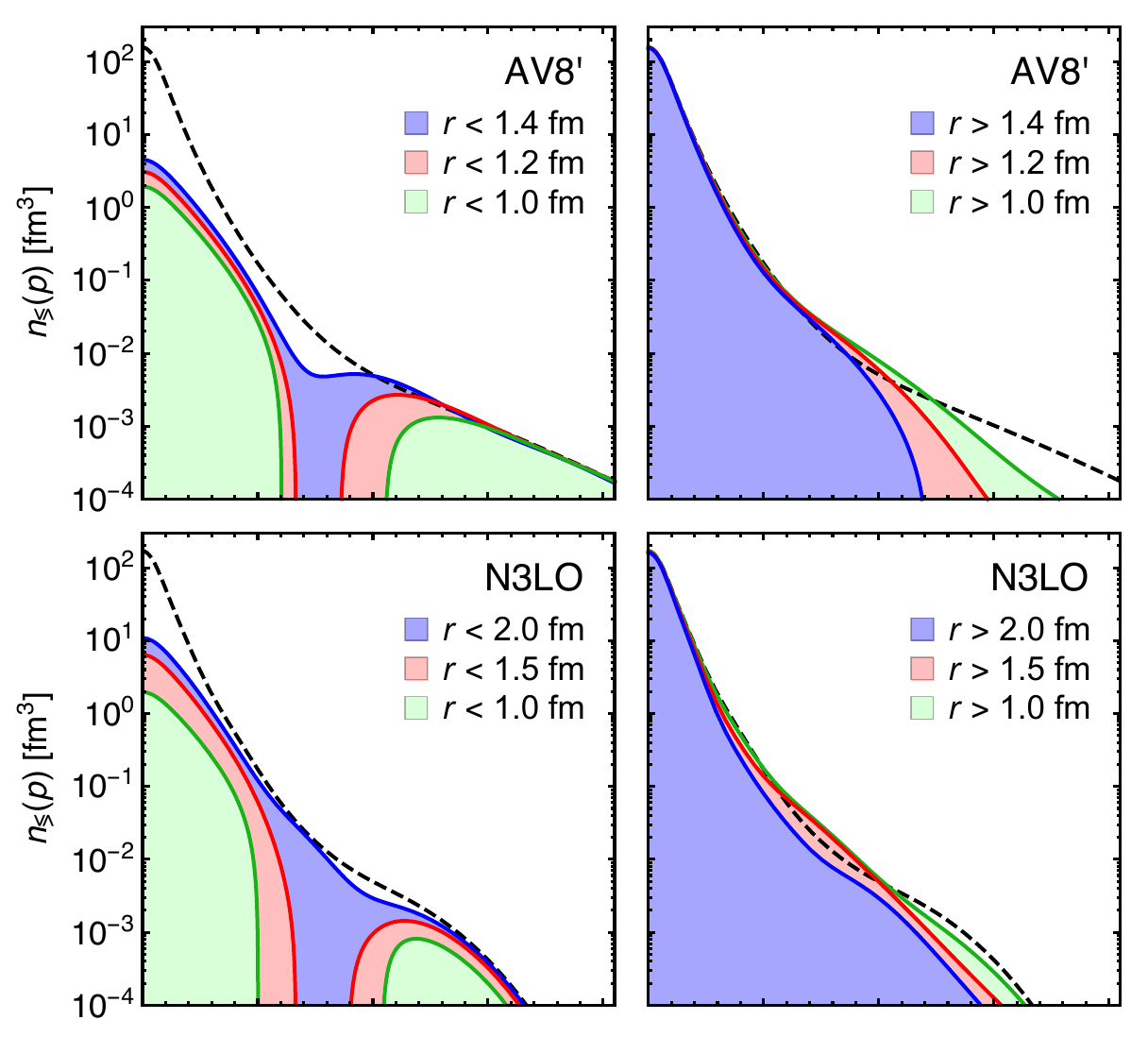}
	\caption{(Color online) Partial momentum distributions for the AV8' (top) and N3LO (bottom) interactions obtained by integrating the Wigner function $W(r,p)$ over small distances (left) and over large distances (right). The dashed line shows the full momentum distribution obtained by integrating over all distances.}
	\label{fig:momentumsep}
\end{figure}

In the upper part of Fig.~\ref{fig:momentumsep} we show the partial momentum distributions 
\begin{equation}
	n_\lessgtr(p) = \int_{r \lessgtr r_\mathit{sep}} dr \: r^2 W(r,p)
\end{equation}
for the AV8' interaction that are obtained by integrating the Wigner function $W(r,p)$ over a region of small distances and large distances respectively. The separation distance distance $r_\mathit{sep}$ was chosen here to be at 1.0, 1.2 and 1.4 fm. If we look at the momentum distributions obtained from the pairs at small distances we find that they are essentially identical with the full momentum distribution for momenta larger than about 2.0, 2.5 and 3.0~$\fm^{-1}$ respectively. They also contribute at smaller momenta but here the large distance pairs dominate. In the mid-momentum region the contribution from small distance pairs is even negative. This is again reflecting the non-classical nature of the Wigner function. The negative contributions are of course compensated by the large distance pairs whose partial momentum distribution in the mid-momentum region is larger than the full momentum distribution. The low-momentum region of the momentum distribution is given almost exclusively by the large distance pairs. 

Interestingly the corresponding partial momentum distributions for the N3LO interaction displayed in the lower part of Fig.~\ref{fig:momentumsep} do not show this nice separation between small and large distances. The separation distances are chosen as 1.0, 1.5 and 2.0 fm. As can be seen one has to go to much larger distances than for the AV8' interaction in order to capture the high momentum components. This can be related to the properties of the non-local regulator that affects the deuteron wave function even at large distances. This is also a reminder that high momenta are related to large curvature of the wave function and not necessarily to short distances. For the AV8' interaction large curvatures are restricted to short distances, at larger distances the wave function is smooth and shows no large curvature. Here short distance and large momenta correlate. This is not the case for the N3LO interaction. The `kinks' generated by the chosen regularization procedure, that can be seen in the deuteron wave function at large distances (Fig.~\ref{fig:wavefunctions}), contribute to the high momentum components of the momentum distribution.

\begin{figure}
	\includegraphics[width=\columnwidth]{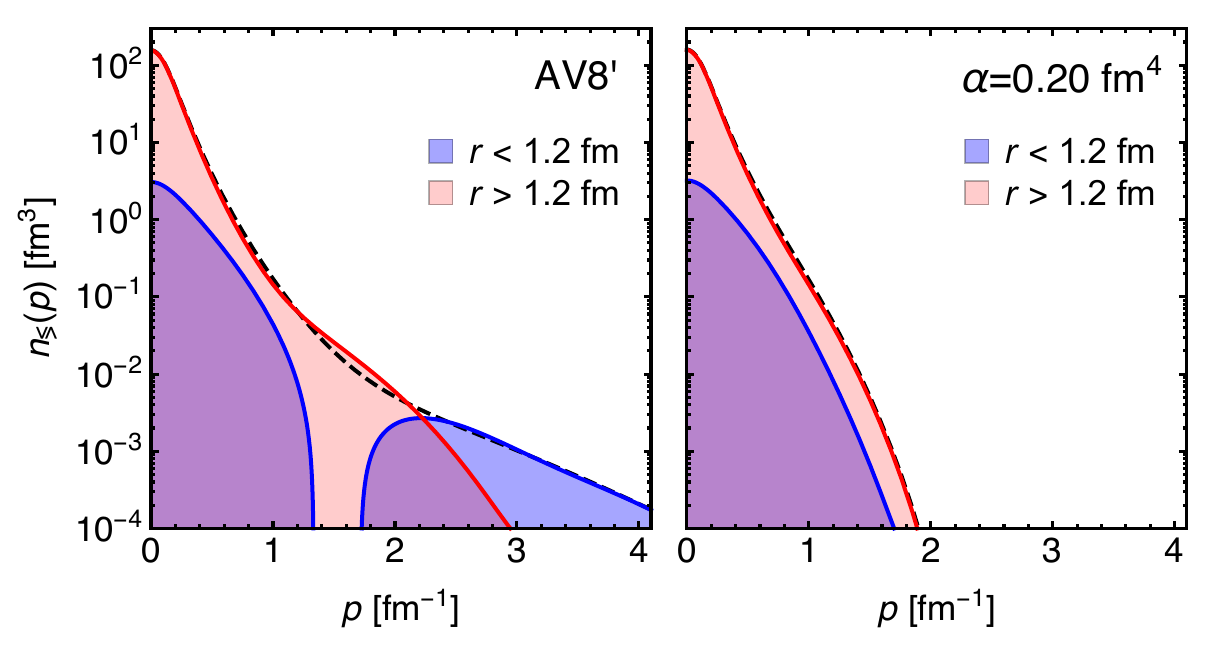}
	\caption{(Color online) Partial momentum distributions for the bare (left) and SRG evolved AV8' interaction (right) obtained by integrating the Wigner function $W(r,p)$ over small distances (blue) and over large distances (red). The dashed line shows the full momentum distribution obtained by integrating over all distances.}
	\label{fig:momentumsep-av8p-srg}
\end{figure}

It is also interesting to compare the partial momentum distributions obtained with the bare interaction with those obtained with a soft SRG evolved interaction. In Fig.~\ref{fig:momentumsep-av8p-srg} this is shown for the bare and SRG evolved AV8' interaction and for a separation distance of 1.2~fm. In case of the soft SRG interaction the momentum distribution has no high-momentum components and the contributions from distances larger than 1.2~fm completely dominate over the contributions from small distances for all momenta.

\subsection{Partial coordinate space distributions $\rho_<(r)$ and $\rho_>(r)$}
\label{sec:coordinatesep}

\begin{figure}
	\includegraphics[width=\columnwidth]{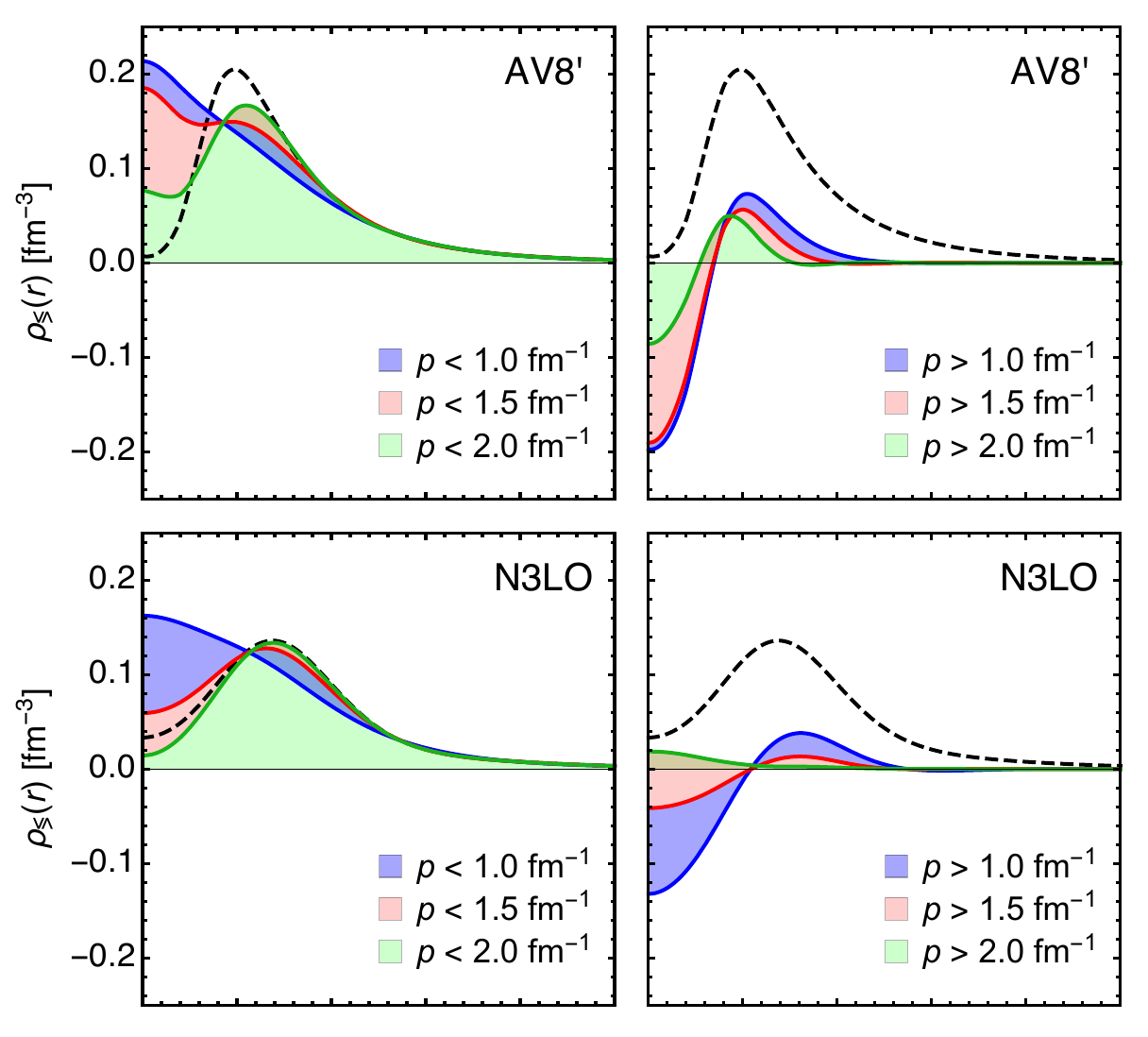}
	\caption{(Color online) Partial coordinate space distributions for the AV8' (top) and N3LO (bottom) interactions obtained by integrating the Wigner function $W(r,p)$ over small momenta (left) and over large momenta (right). The dashed line shows the full distribution obtained by integrating over all momenta.}
	\label{fig:coordinatesep}
\end{figure}

In the upper part of Fig.~\ref{fig:coordinatesep} we show the partial coordinate space densities
\begin{equation}
	\rho_\lessgtr(r) = \int_{p \lessgtr p_\mathit{sep}} dp \: p^2 W(r,p)
\end{equation}
obtained from low- and high-momentum pairs for the AV8' interaction. The separation momentum $p_\mathit{sep}$ is taken to be at 1.0, 1.5 and 2.0~$\fm^{-1}$. Here we can clearly see that the coordinate space distribution at large distances $r \gtrsim 2\:\fm$ is determined exclusively by low-momentum pairs. It is interesting to observe that the low momentum pairs do not show a suppression of the coordinate space density at small distances. The partial coordinate space density $\rho_<(r)$ from the low-momentum pairs looks like what one would expect for a smooth potential without a repulsive core which does not induce strong correlations and thus would be suited for mean-field descriptions. The partial coordinate space density $\rho_>(r)$ density obtained from the high-momentum pairs shows two characteristic features (right hand side of Fig.~\ref{fig:coordinatesep}). The first observation is that it is restricted to small distances and the second is that it takes negative values at small distances. In this picture one understands the correlation hole in the full coordinate space density as the cancellation of contributions from low- and high-momentum pairs with different signs. This genuine feature of the Wigner representation is illustrated and discussed further in Appendix~\ref{app:schematic}.

For the N3LO interaction the partial coordinate space densities shown in lower part of Fig.~\ref{fig:coordinatesep} are less sensitive to very high momenta and the total momentum distribution is close to that obtained by taking all momentum up to 2~$\fm^{-1}$ into account. 

\begin{figure}
	\includegraphics[width=\columnwidth]{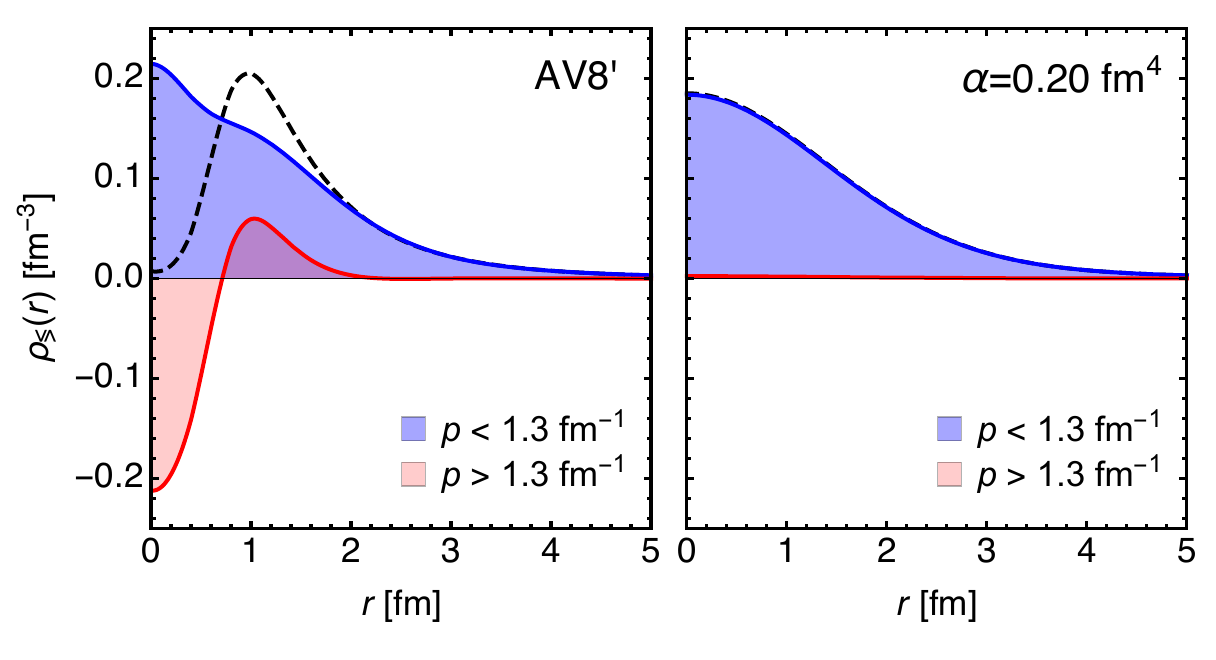}
	\caption{(Color online) Partial coordinate space distributions obtained by integrating the Wigner function $W(r,p)$ over ranges of small and large momenta. On the left for the bare AV8' interaction on the right for the SRG evolved AV8' interaction ($\alpha {=} 0.20\:\fm^4$) The dashed lines shows the full distributions obtained by integrating over all momenta.}
	\label{fig:coordinatesep-srg}
\end{figure}

Again it is interesting to isolate the contribution from short-range correlations. In Fig.~\ref{fig:coordinatesep-srg} we compare the partial coordinate space densities $\rho_\lessgtr(r)$ for the bare AV8' and the SRG evolved AV8' interactions. For the soft SRG interaction the coordinate space density is determined completely by relative momenta up to $p \approx 1.3\:\fm^{-1}$. With the soft interaction the highest density is found when the nucleons sit on top of each other. As already discussed the situation is very different for the bare interaction. Here we can clearly divide the contributions from low momenta that provide a partial density distribution that looks similar to that of the soft interaction and from high momenta that contribute only at distances up to $r \approx 2\:\fm$ and that become strongly negative at small distances. The effect of reducing short-range correlations by softening the interaction with the SRG evolution is also highly visible in the Wigner representation.

\subsection{Angular correlations $W(r,p,\vartheta)$}

\begin{figure*}
	\includegraphics[width=0.31\textwidth]{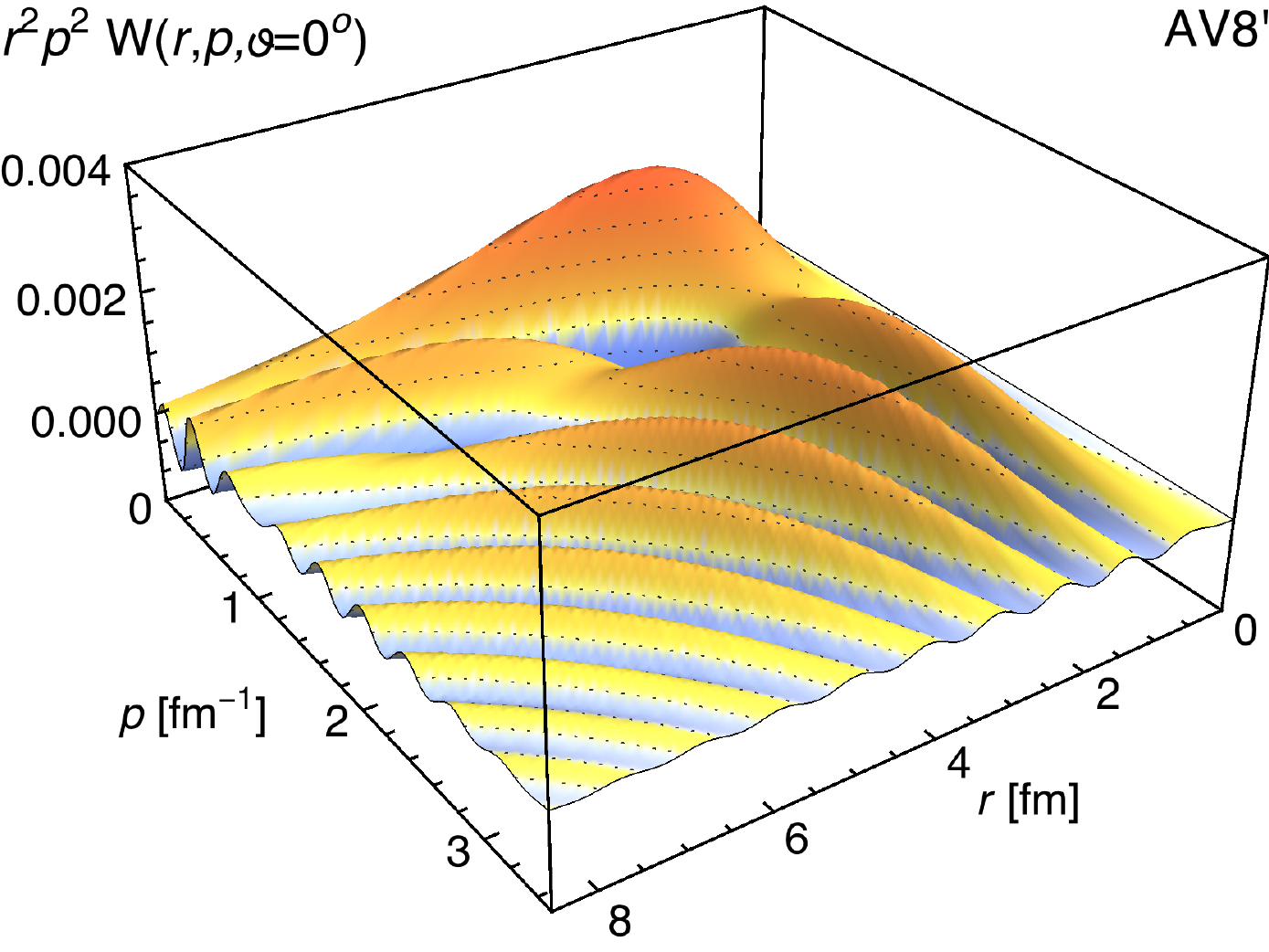}\hfil
	\includegraphics[width=0.31\textwidth]{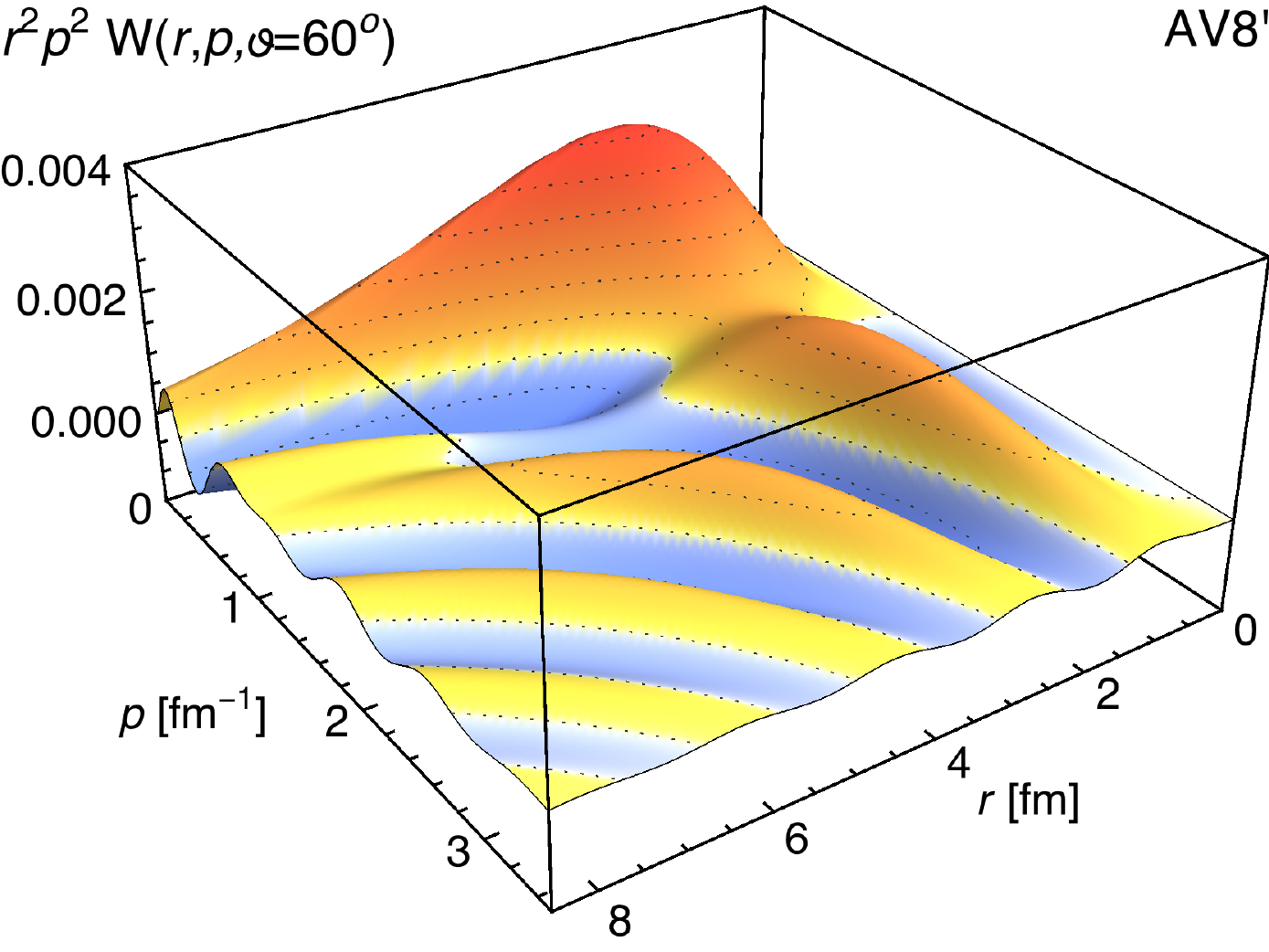}\hfil
	\includegraphics[width=0.31\textwidth]{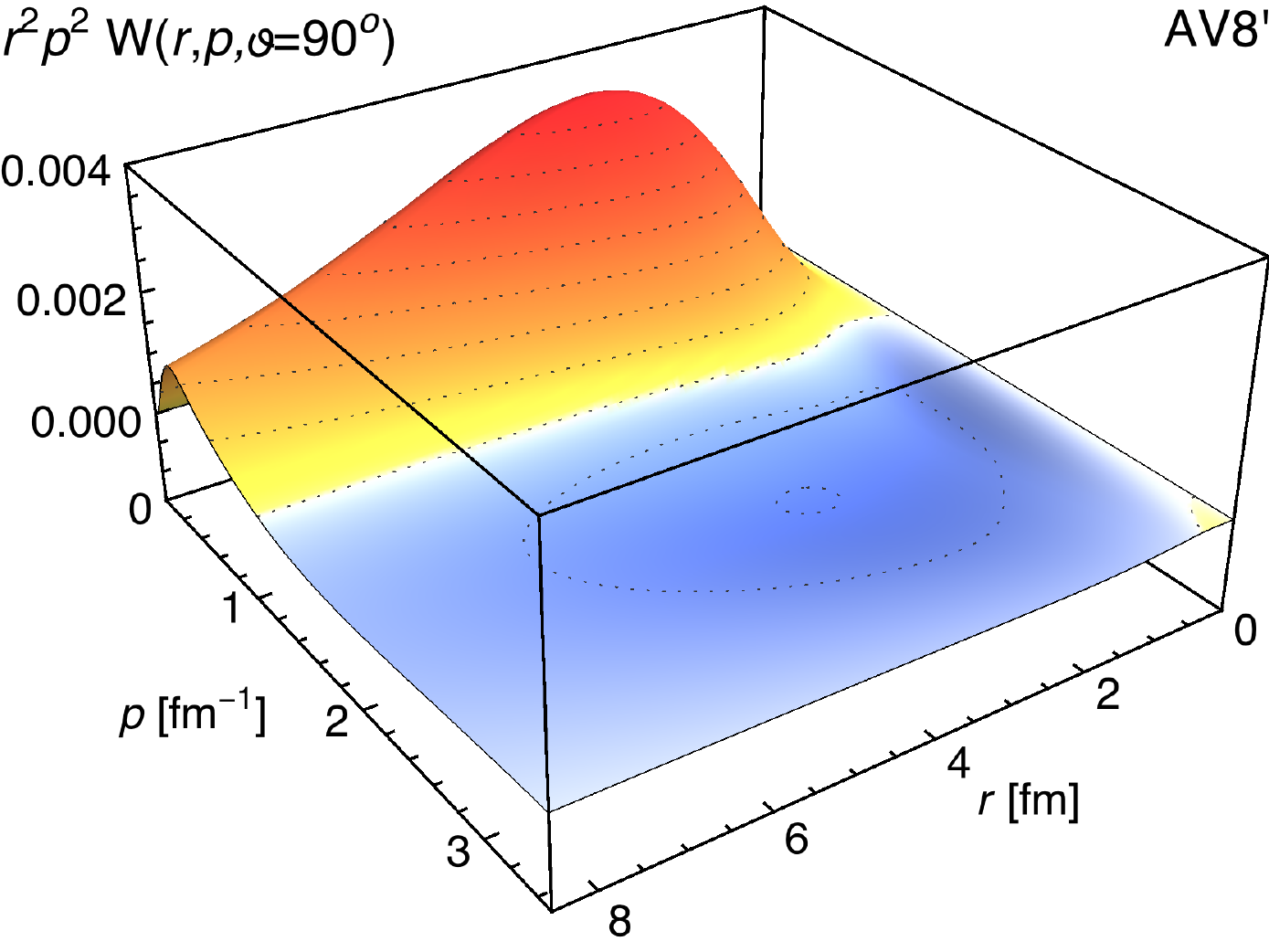}
	\caption{(Color online) Wigner function of the deuteron for the AV8' interaction scaled with phase space volume element $r^2 p^2 W(r,p,\vartheta)$ for $\vartheta=0^\circ$ (left), $\vartheta=60^\circ$ (middle) and $\vartheta=90^\circ$ (right).}
	\label{fig:wignerscaledrptheta-av8p}
\end{figure*}

\begin{figure*}
	\includegraphics[width=0.31\textwidth]{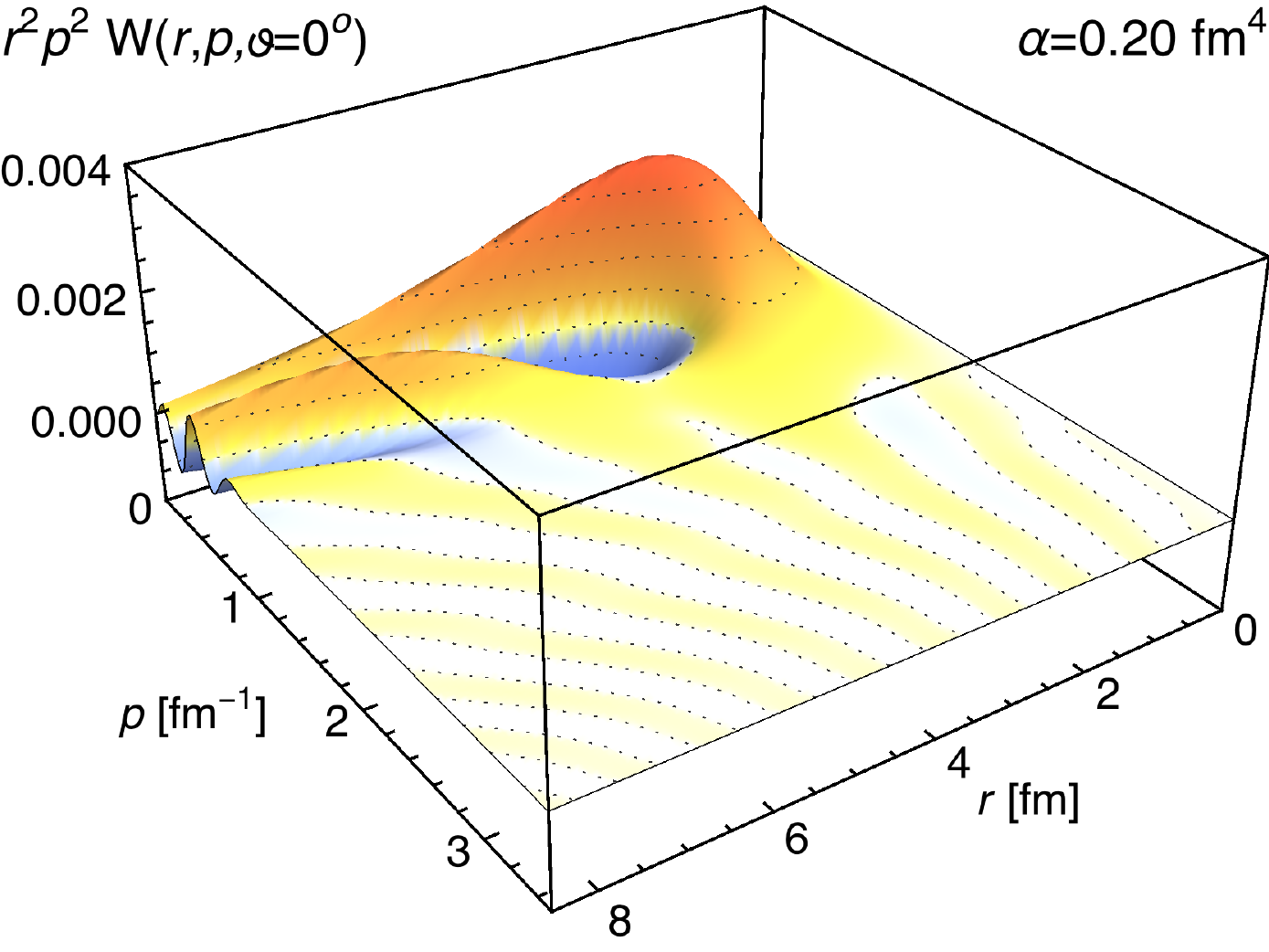}\hfil
	\includegraphics[width=0.31\textwidth]{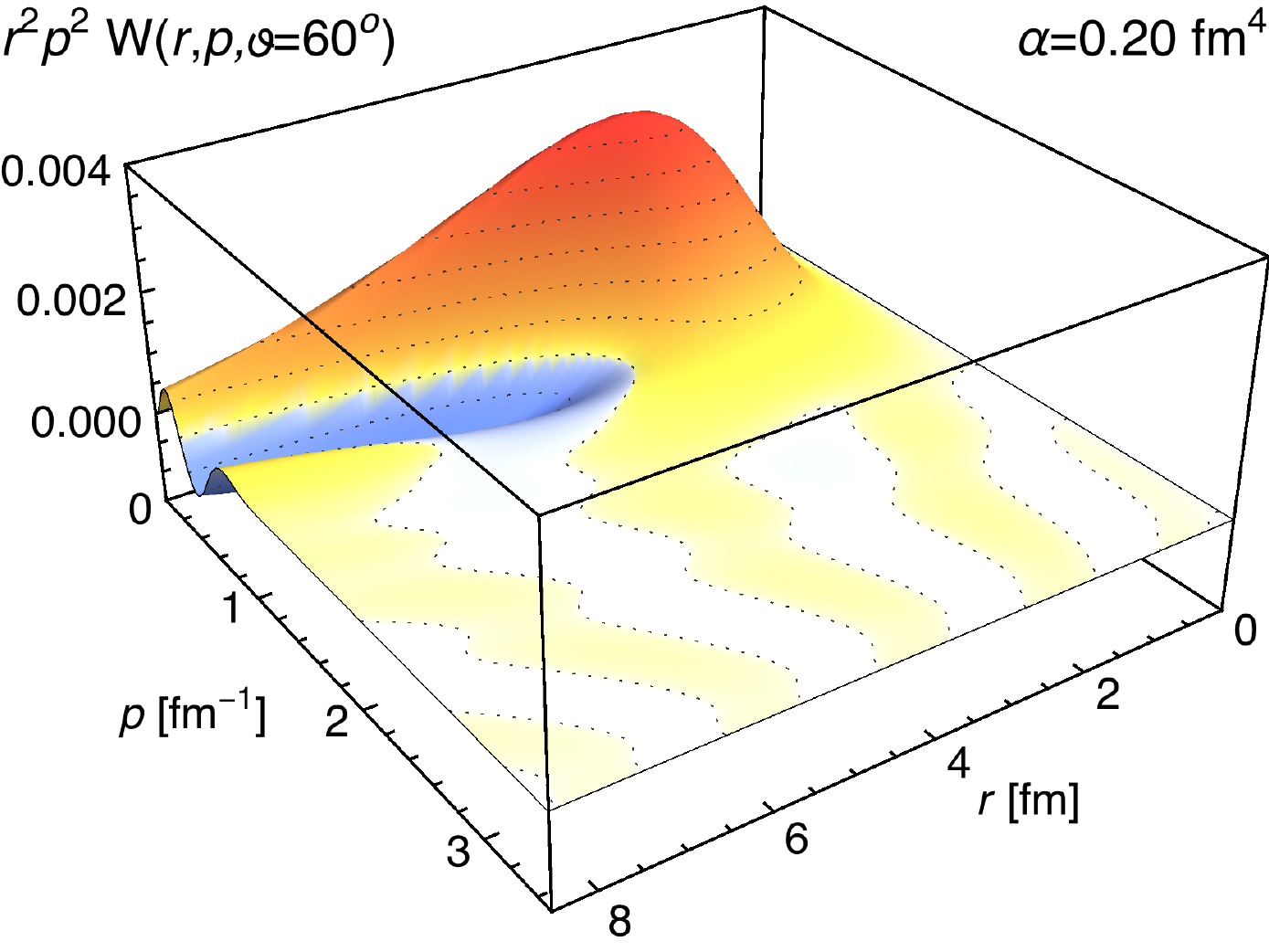}\hfil
	\includegraphics[width=0.31\textwidth]{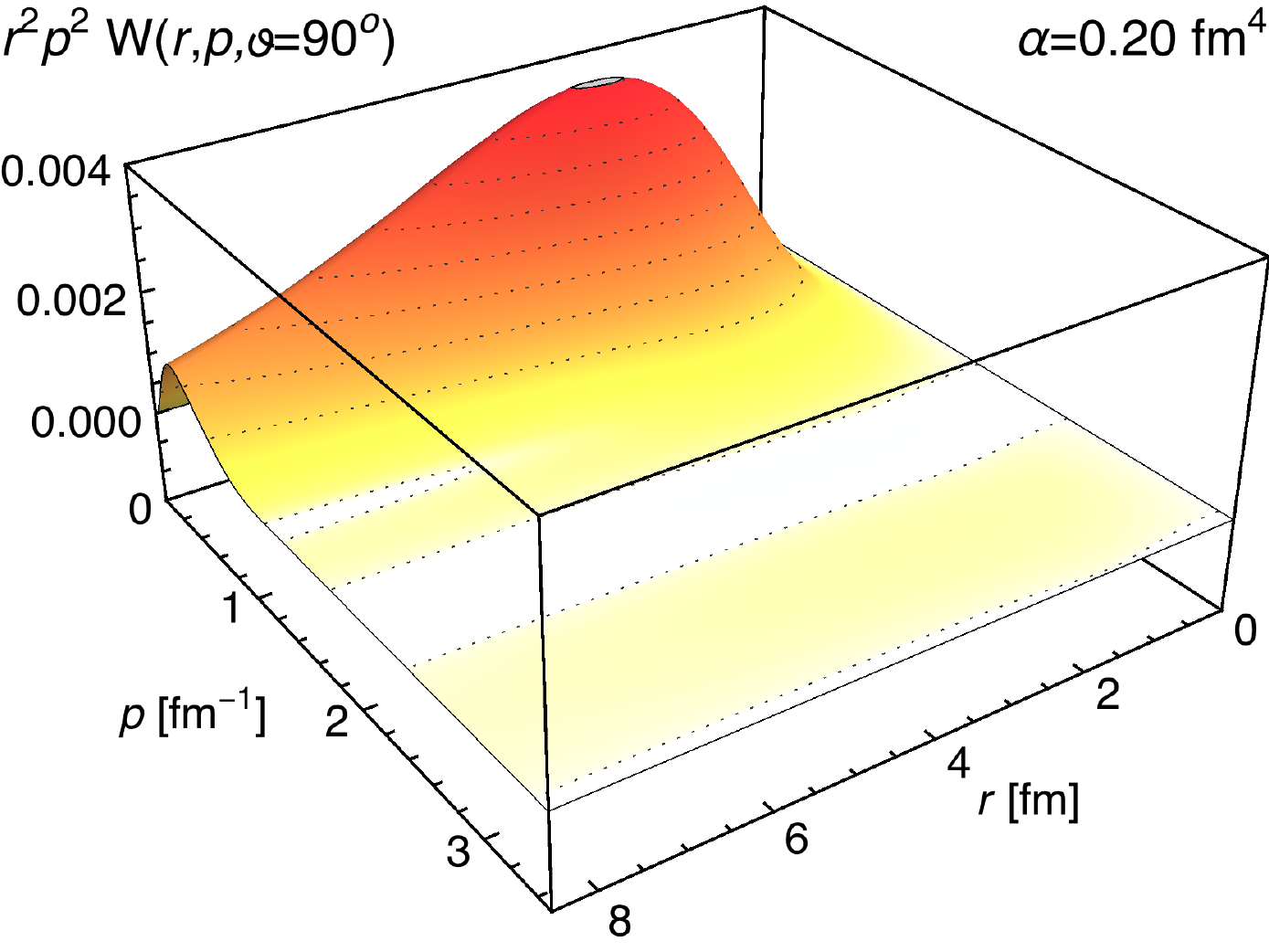}
	\caption{(Color online) Wigner function of the deuteron for the SRG evolved AV8' interaction ($\alpha {=} 0.20\:\fm^4$) scaled with phase space volume element $r^2 p^2 W(r,p,\vartheta)$ for $\vartheta=0^\circ$ (left), $\vartheta=60^\circ$ (middle) and $\vartheta=90^\circ$ (right).}
	\label{fig:wignerscaledrptheta-av8p_srg2000}
\end{figure*}

In this section we explore the dependence of the Wigner function $W(\vec{r},\vec{p}$) on the orientations of the distance vector $\vec{r}$ and the relative momentum vector $\vec{p}$. Due to rotational invariance the Wigner function depends only on the relative orientation of $\vec{r}$ and $\vec{p}$, so that $W(\vec{r},\vec{p}) = W(r,p,\vartheta)$.

This dependence on the angle $\vartheta$ is shown in Fig.~\ref{fig:wignerscaledrptheta-av8p} for the AV8' interaction. The oscillatory behavior of the Wigner function depends strongly on the angle $\vartheta$. The uncertainty relation demands that distance and relative momentum in the same direction must fulfill
\begin{equation}
	\Delta x_i \cdot \Delta p_i \geq 1/2 \: .
\end{equation}
However no such relation holds for distance vectors and momenta perpendicular to each other. In this case ($\vartheta = 90^\circ$) the Wigner function is very smooth and is concentrated in the low-momentum region. If the vectors are parallel or antiparallel ($\vartheta = 0^\circ$) the Wigner function oscillates strongly. Compared to the perpendicular case we can also observe that the Wigner function is reduced in the low-momentum region but we find relatively large contributions at higher momenta. 

\begin{figure}
	\includegraphics[width=\columnwidth]{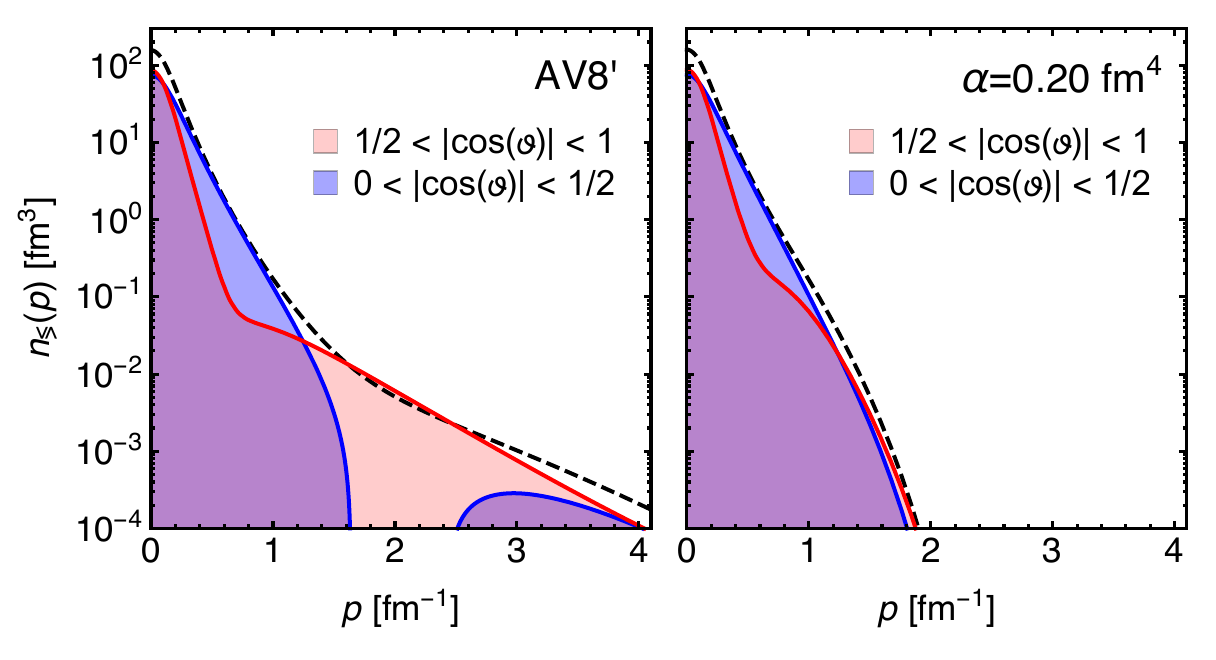}\\
	\includegraphics[width=\columnwidth]{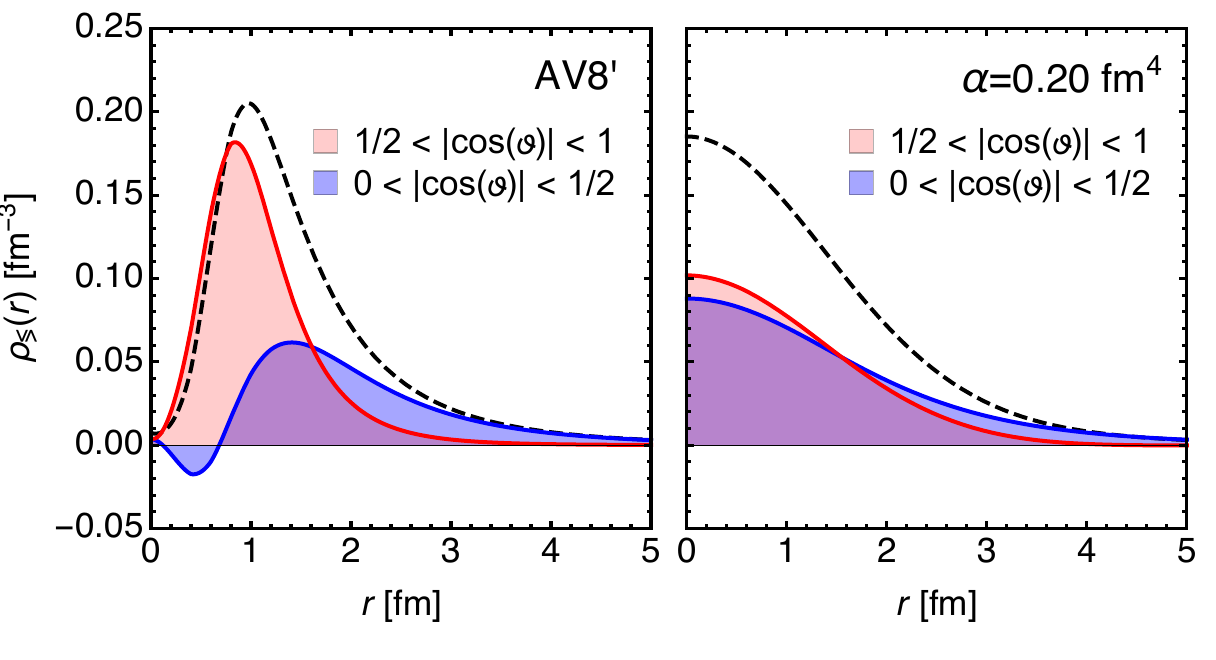}
	\caption{(Color online) Partial momentum (top) and coordinate space (bottom) distributions obtained by integrating over small and large angles respectively. For the bare (left) and the SRG evolved AV8' (right) interaction.}
	\label{fig:momentum-coordinate-theta-av8p-srg}
\end{figure}

This indicates that the low-momentum part is dominated by the contributions where distance vectors and relative momenta are mostly perpendicular whereas the high-momentum or short-range component is related to configurations when distance and relative momentum are parallel or antiparallel. This is highlighted in Fig.~\ref{fig:momentum-coordinate-theta-av8p-srg} where we show the partial momentum (coordinate space) densities obtained by integrating over all distances (momenta) but restrict the integration range for the angles. The coordinate space distribution at large distances is given by the contributions from large angles whereas the short distances are dominated by the small angle contributions. Interestingly this dominance of small angle contributions at short distances is not seen for the SRG evolved interaction. Here the long-range part is still dominated by large angle contributions but at short and mid distances small and large angles contribute almost equally. In the partial momentum distributions we can observe that the large angle contributions dominate the low momentum region with the exception of very small momenta where small and large angles contribute equally. The mid and high momentum region on the other hand is dominated by the small angle contributions.

Fig.~\ref{fig:wignerscaledrptheta-av8p_srg2000} displays the Wigner functions at the same angles as Fig.~\ref{fig:wignerscaledrptheta-av8p} but for the SRG evolved AV8' interaction. Again for $\vec{r}$ perpendicular to $\vec{p}$ ($\vartheta = 90^\circ$) the Wigner function is very smooth with a low momentum peak that is very similar to that obtained with the bare interaction. At smaller angles one can observe again oscillatory behavior that is however strongly suppressed in amplitude at larger momenta compared to the results with the bare interaction.

\subsection{Angle distribution $\Theta(\vartheta)$}

\begin{figure}
	\centering
	\includegraphics[width=\columnwidth]{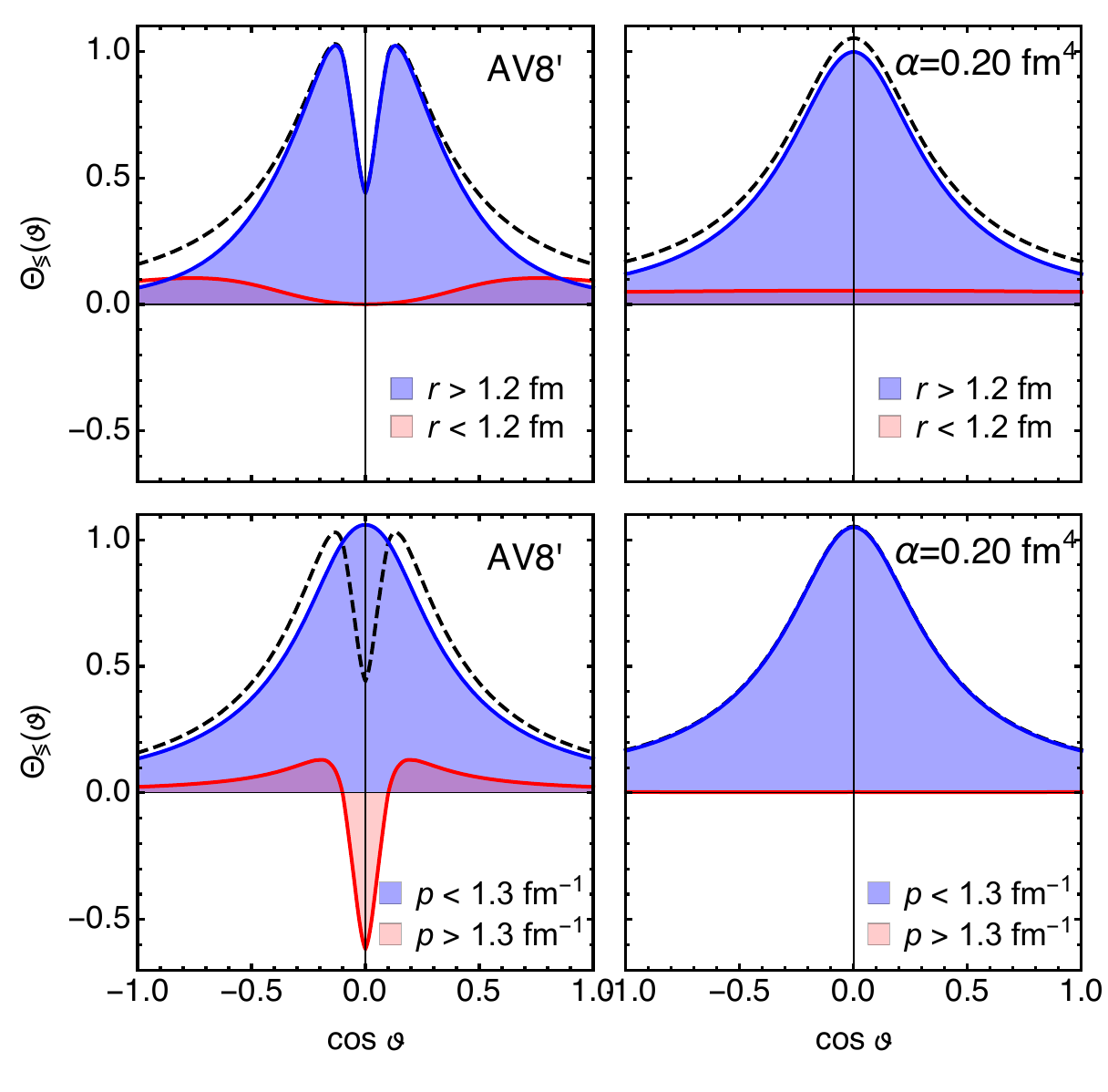}
	\caption{(Color online) $\Theta_\lessgtr(\vartheta)$ for the AV8' (left) and the SRG evolved AV8' interaction (right). Disentangling contributions from small and large distances (top) and low and high momenta (bottom). The dashed line shows the full angle distribution $\Theta(\vartheta)$.}
	\label{fig:angle}
\end{figure}

Inspired by the discussion for the Hydrogen atom in \cite{dahl82b} we integrate the Wigner function $W(r,p,\vartheta)$ over distances and momenta to obtain the probability as a function of the angle between distance and relative momentum vector
\begin{equation}
	\Theta(\vartheta) = 8 \pi^2 \int dr \: r^2 \int dp \: p^2 \: W(r,p,\vartheta) \: .
\end{equation}
Also here we can define partial angle distributions $\Theta_\lessgtr(\vartheta)$ by integrating only over small or large distances or over low and high momenta. The results are shown in Fig.~\ref{fig:angle} for the AV8' and SRG evolved AV8' interactions. For the soft SRG interaction the probability peaks at $\vartheta = 90^\circ$. For a classical circular orbit only the angle $\vartheta = 90^\circ$ would contribute. Another extrem case is given by a simple Gaussian wave function which has no correlation between the orientation of distance vector and relative momentum so that $\Theta(\vartheta)$ is flat. The Wigner function obtained with the soft SRG interaction however deviates from the Gaussian wave function and shows a correlation in the relative orientation of distance vector and relative momentum that is qualitatively similar to that found in the hydrogen atom \cite{dahl82b}. For the bare interaction the behavior of $\Theta(\vartheta)$ is more complicated. There is a pronounced dip at $\vartheta = 90^\circ$ that is obviously related to the high momentum pairs.

\subsection{Spin dependence and tensor correlations $W_{M_S,M_S}(\vec{r},\vec{p})$}

\begin{figure}
	\centering
	\includegraphics[width=0.33\textwidth]{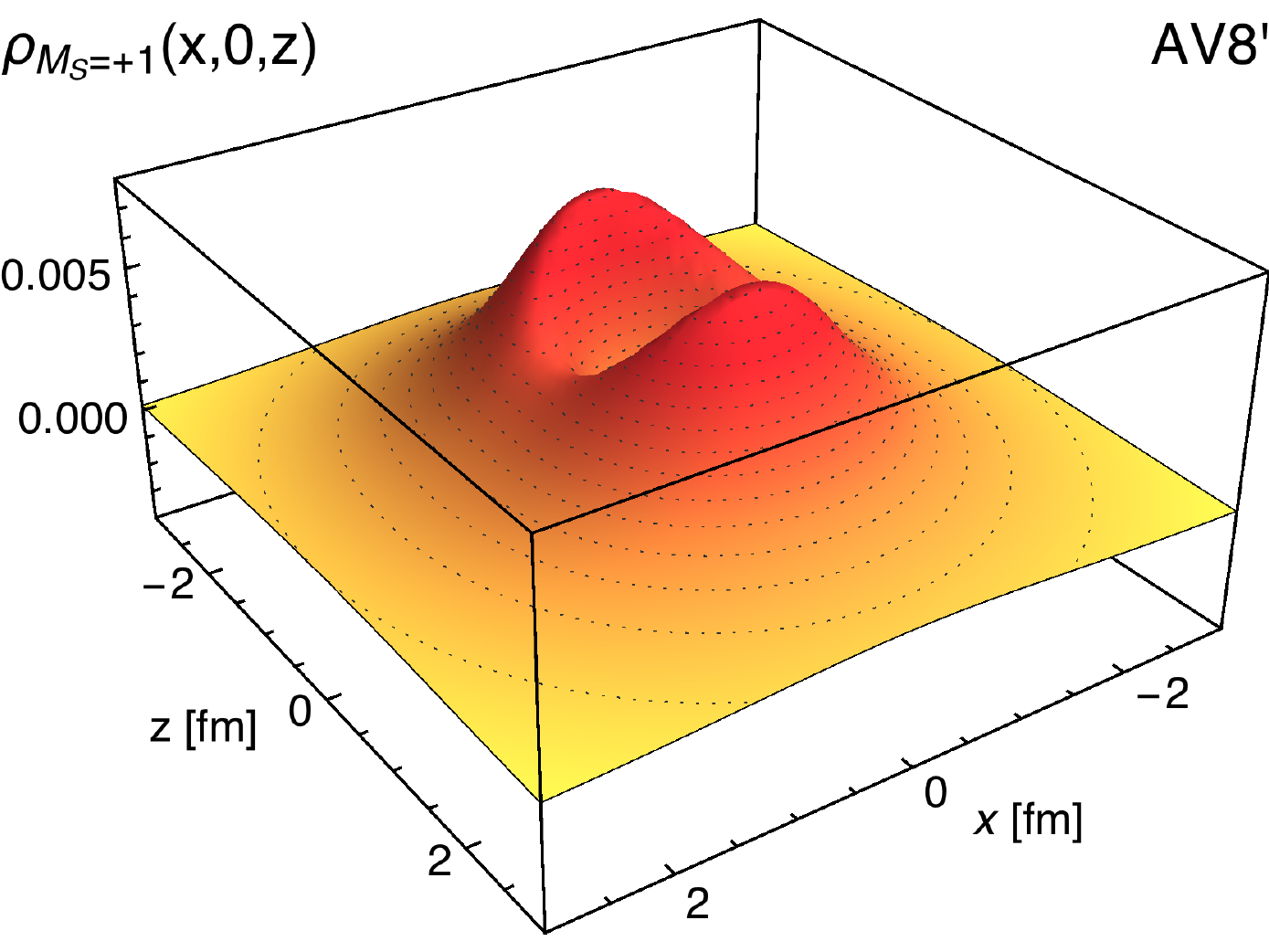}\\
	\includegraphics[width=0.33\textwidth]{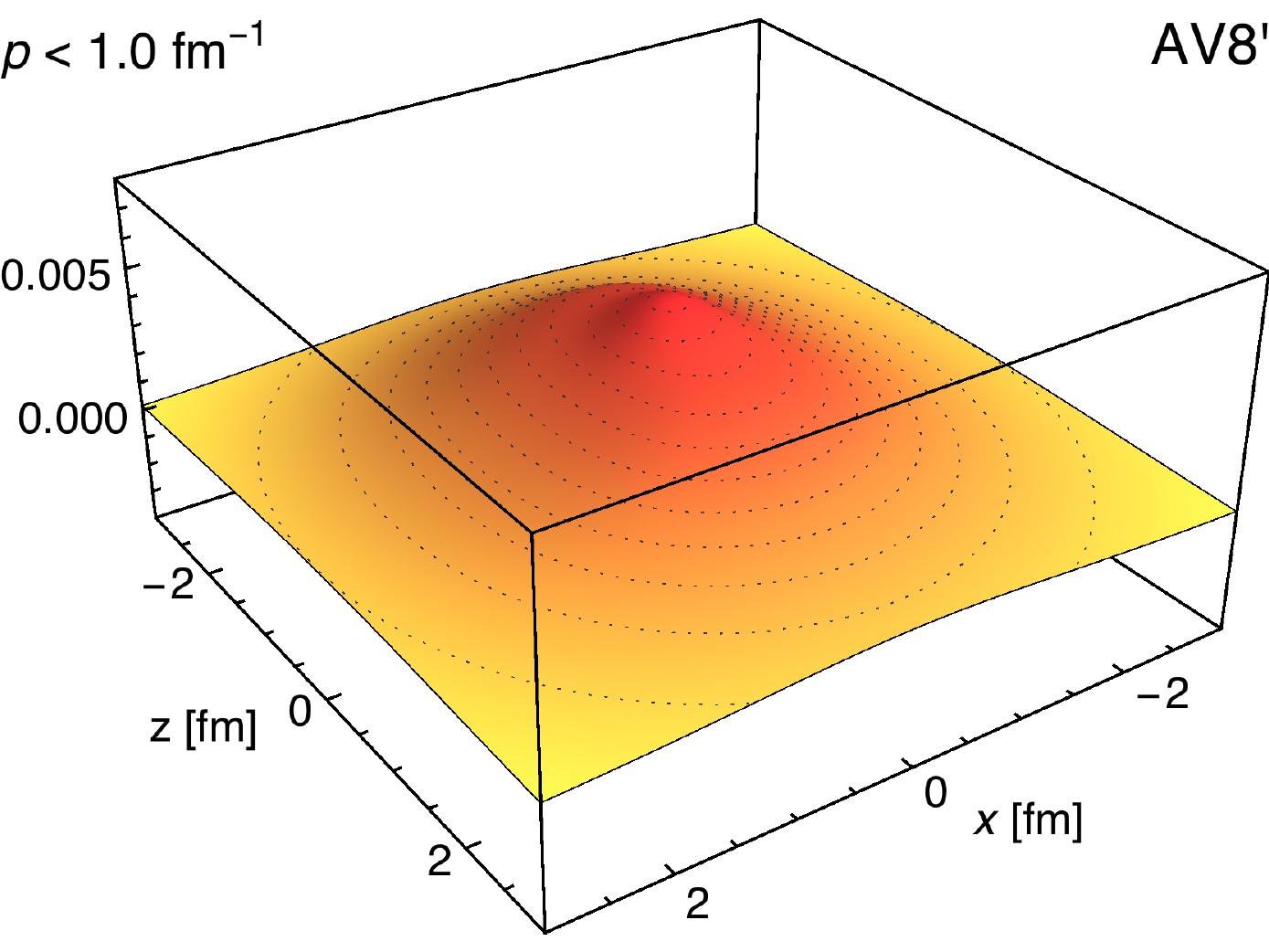}\\
	\includegraphics[width=0.33\textwidth]{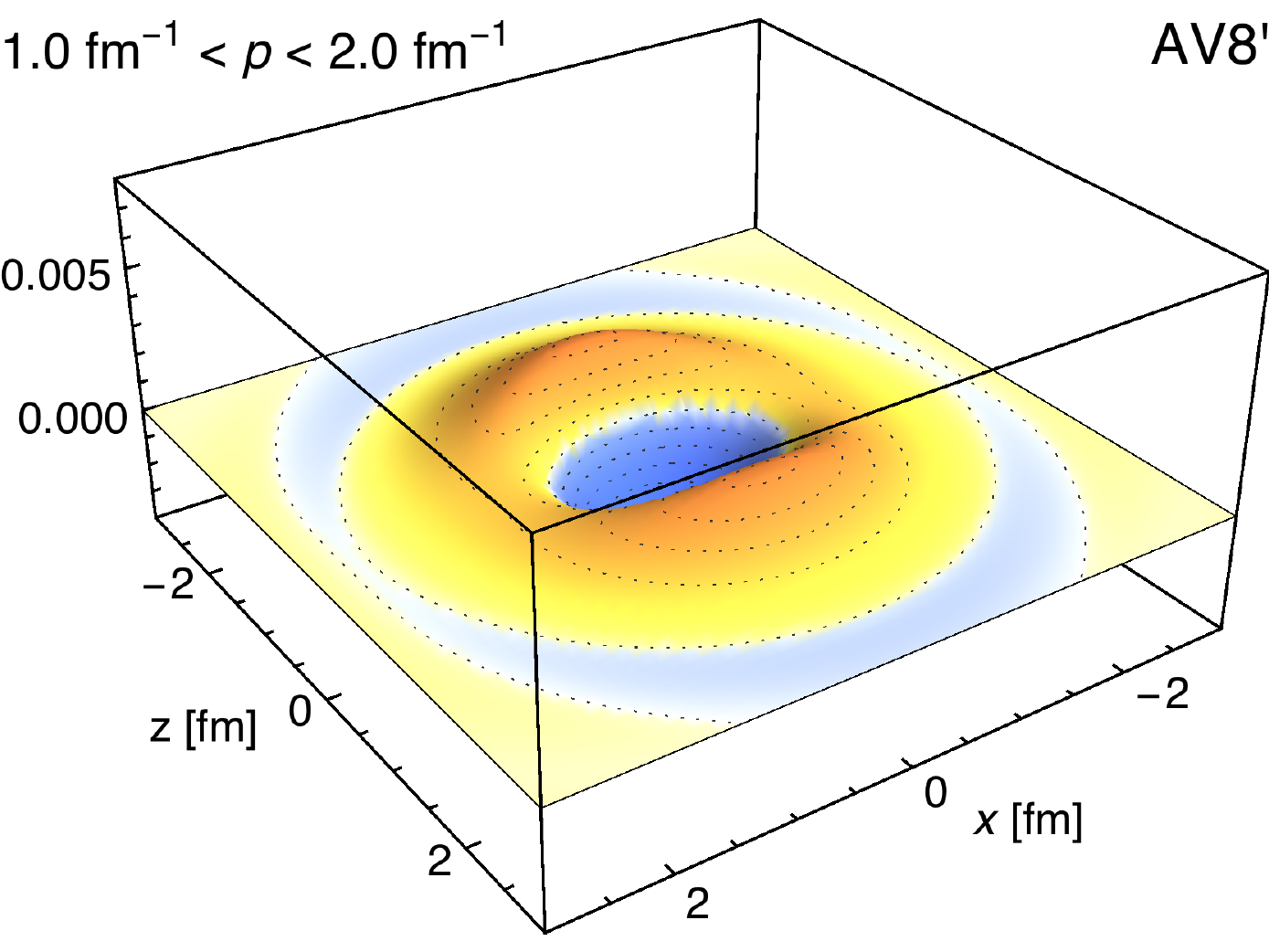}\\
	\includegraphics[width=0.33\textwidth]{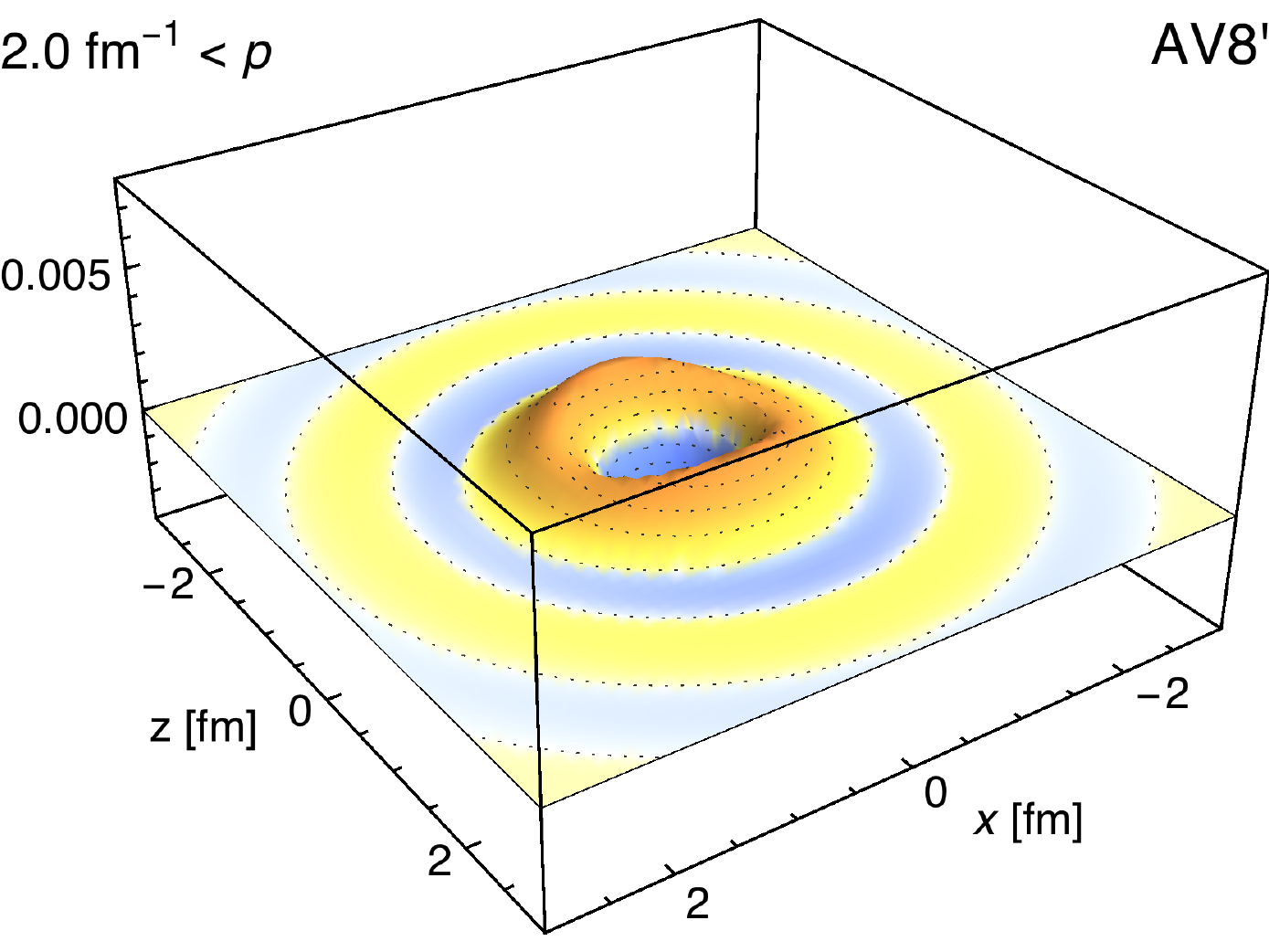}
	\caption{(Color online) Total (top) and partial coordinate space densities $\rho_{M_S=+1}(x,y {=} 0,z)$ for the deuteron with the AV8' interaction. The spin is aligned along the $z$-axis. Total density on top, followed by the partial densities from low, medium, and high momentum regions.}
	\label{fig:rhoxzpspin}
\end{figure}

Although the Wigner function $W_{M_S,M_S}(\vec{r},\vec{p})$ is a scalar quantity as we have averaged in Eq.~\eqref{eq:densitymatrix} over all orientations of the total angular momentum it is possible to have correlations between the spin orientation given by $M_S$ and the spatial orientations given by $\vec{r}$ and $\vec{p}$. Such correlations indeed exist in the deuteron because of the tensor force and the admixture of the $L=2$ component in the wave function. In the case of the coordinate space densities this leads to the famous ``dumpbell'' and ``donut'' shapes \cite{schiavilla07,ucom03,src11} for $M_S=\pm1$ and $M_S=0$ respectively. Using the Wigner functions $W_{M_S,M_S}(\vec{r},\vec{p}$) we can try to shed further light on the origin of these tensor correlations.

In Fig.~\ref{fig:rhoxzpspin} where we show the total density distribution $\rho_{M_S}(\vec{r})$ and the partial density density distributions
\begin{equation}
  \rho_{M_S}(\vec{r} ; p_l {<} |\vec{p}| {<} p_h) = \int_{p_l<|\vec{p}|}^{|\vec{p}|<p_h} d^3p \: W_{M_S,M_S}(\vec{r},\vec{p}) \: .
\end{equation}
The total density distribution $\rho_{M_S=1}(\vec{r})$ illustrates the correlations in the deuteron quite nicely. At small distances the density is suppressed due to the repulsive core of the interaction. As the tensor interaction is attractive when the nucleons are aligned along the spin orientation the density distribution is also aligned along the spin direction (for $M_S=1$ along the $z$-axis). In the perpendicular direction the density is suppressed. 

The partial density distributions allow to analyze the contributions from particular momentum regions to the total density. The contributions from small momenta $p \lesssim 1\:\fm^{-1}$ show almost no correlation between spin and density. As discussed in Sec.~\ref{sec:coordinatesep} for the spin independent partial coordinate space densities the low momentum region also show no suppression of the density at small distances. The strongest correlations with the spin direction can be seen in the contributions from the mid-momentum region. This is also characterized by negative contributions to the density in the region of the repulsive core and a stronger spatial localization. The contributions from higher momenta show less tensor correlations and are more localized.

\begin{figure}
	\centering
	\includegraphics[width=0.33\textwidth]{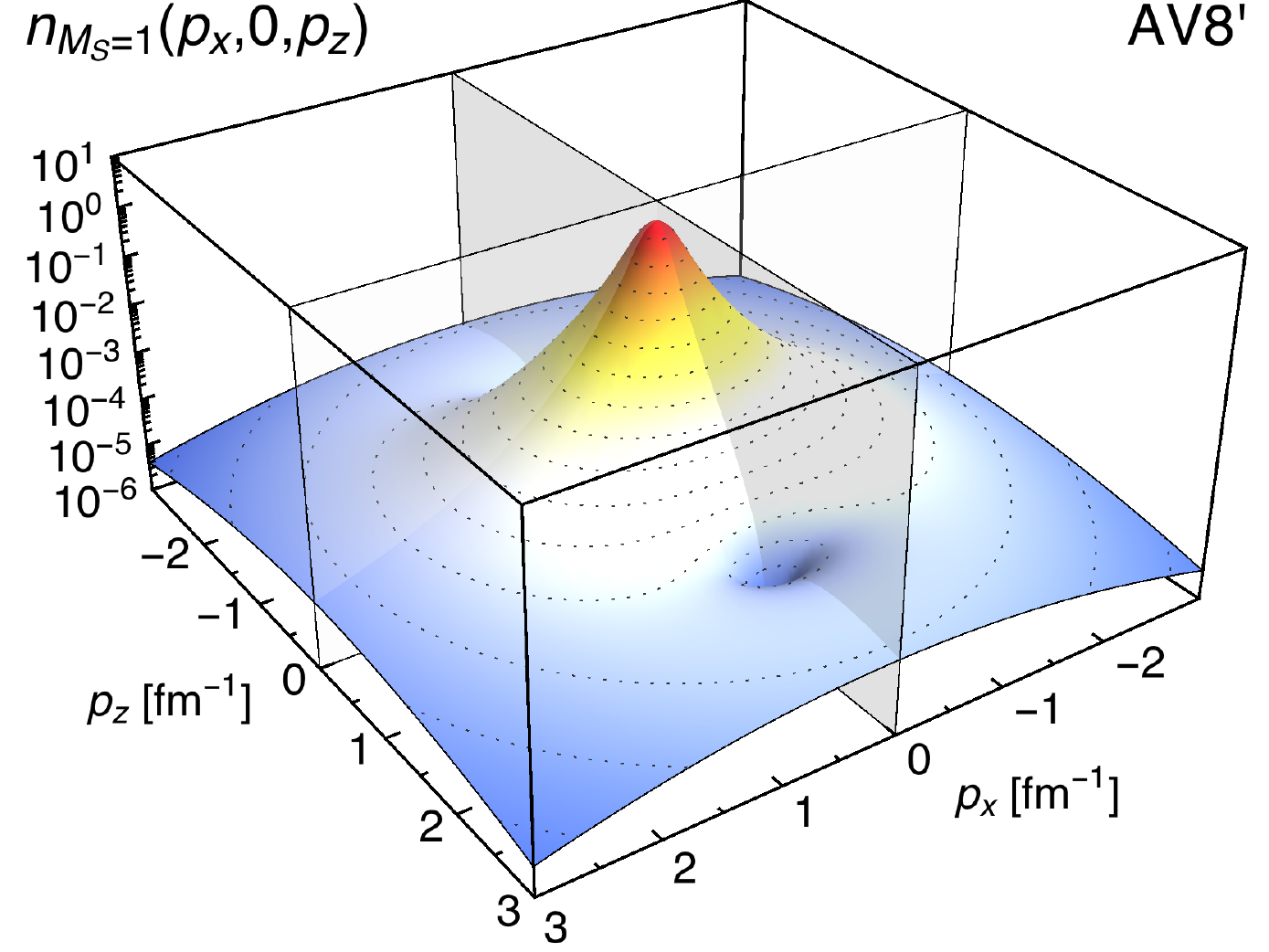}
	\caption{(Color online) Momentum space density $n_{M_S=1}(p_x,p_y {=} 0,p_z)$ for deuteron spin aligned along the $z$-axis $M_S=1$.}
	\label{fig:nxzpspin}
\end{figure}

The total momentum distribution $n_{M_S=1}(\vec{p})$ displayed in Fig.~\ref{fig:nxzpspin}, shows a clear dependence on the spin orientation in the mid-momentum region. The momentum distribution has a dip at $p_z \approx \pm 1.5\:\fm^{-1}$ for momenta parallel to the spin orientation but takes finite values for larger angles between spin and momentum direction.

\section{Summary and Conclusions}
\label{sec:summary}

The Wigner function can be considered as the phase space representation of quantum mechanics and allows to study correlations between coordinate and momentum space. In this paper we restrict ourselves to a very simple system -- the deuteron -- to investigate short-range correlations with the Wigner function. Despite this restriction to the two-body system the results are expected to be representative also for heavier nuclei due to the universality of short-range correlations in nuclei \cite{src11,alvioli12}. Short-range correlations have been studied either in coordinate space, where a suppression of the density at small distances induced by the repulsive core of the nucleon nucleon interaction is observed, or in momentum space where short-range correlations are reflected in high-momentum components of the one-, or more clearly, of the two-nucleon momentum distributions. The Wigner function furthermore makes a comparison with the classical picture possible and highlights the quantum-mechanical nature of short-range correlations.

The full information about the system is contained in the Wigner function $W_{M_S,M_S'}(\vec{r},\vec{p})$ that depends not only on the absolute values of the distance $r$ and the relative momentum $p$ of the nucleons but also on the orientation of  distance and momentum with respect to each other and with respect to the orientation of the spin $\vec{S}$. Reduced Wigner functions are obtained by summing over all possible spin orientations, $W(\vec{r},\vec{p})$, and by further integrating over the angles between distance and relative momentum, $W(r,p)$. 

The Wigner functions $W(r,p)$ obtained with realistic AV8' and N3LO interactions show a characteristic behavior. One can distinguish between a dominant low-momentum region that is unaffected by short-range correlations and a high-momentum region that shows a shoulder at distances up to 1.5~fm and that extends to high relative momenta. Another characteristic feature is an oscillating behavior of the Wigner function extending to large arguments $r p \approx \mathrm{const.}$. The contribution of short-range correlations is highlighted when the Wigner functions for the bare AV8' and N3LO interactions are compared to those obtained with softened interactions in the SRG approach. The SRG transformation decouples low- and high-momentum modes and essentially eliminates short-range correlations. This is reflected in the Wigner functions that are very similar to those obtained with the bare interactions in the low-momentum region, but that become more and more suppressed in the high-momentum region with increasing flow parameter.

The oscillating behavior is strongly related to the angular dependence of the Wigner function $W(r,p,\vartheta)$. Strong oscillations are found for $\vartheta=0^\circ$ when distance vector and relative momentum are parallel, but there are no oscillations for $\vartheta=90^\circ$ when distance vector and relative momentum are perpendicular. This is a direct consequence of Heisenberg's uncertainty principle. In the reduced Wigner function $W(r,p)$ one averages over all angles and obtains an oscillatory behavior between these two extremes. The general behavior of the Wigner function with its low- and high-momentum part can be understood in a schematic model where the deuteron wave function is described as a superposition of a long-ranged (low-momentum) and a short-ranged (high-momentum) Gaussian. This schematic model illustrates the importance of the interference between the low- and high-momentum components. This interference effect plays an essential role for the oscillatory behavior of the Wigner function, the suppression of the coordinate space density at short distances and the dip in the $S$-wave momentum distribution around 1.5~$\fm^{-1}$. 

If the Wigner function is interpreted as a quasi-probability distribution one can define partial momentum distributions that are obtained by integrating over pairs at small and large distances respectively. In case of the AV8' interaction there is a clear separation between a small- and large distance scale. Pairs at small distances generate the high-momentum components. This separation is not so clear for the N3LO interaction. This interaction contains regulators that affect the wave function even at large distances and generate high-momentum components. The partial coordinate space densities obtained by integrating over low- and high-momentum regions respectively provide a complimentary picture to the momentum distributions. Low-momentum pairs generate coordinate space densities that do not reflect the short-range repulsion and show no suppression at short distances. To generate this suppression at small distances one needs the high-momentum pairs that contribute negatively to the total density at small distances.

Short-range correlations are not only generated by the repulsive core of the interaction but also by the tensor force. The tensor force is responsible for the dominance of proton-neutron over proton-proton pairs as observed for example in $(e,e'pN)$ experiments \cite{subedi08,korover14}. To isolate the effects of tensor correlations it is necessary to look at correlations between the spin orientation and spatial densities. Correlations between spin and spatial degrees of freedom have their origin in the $D$-wave component of the deuteron and these become only visible in the mid-momentum region around 1.5~$\fm^{-1}$ where the $S$-wave component has a node and the $D$-wave component becomes the dominating component. 

Our analysis of the Wigner function emphasizes the fact that short-range correlations are of quantum-mechanical nature and can not be understood in classical terms. In a classical picture the system would occupy a single point in phase space. Due to the repulsive core of the interaction this would correspond to an orbit at a distance of about 1~fm where the potential has a minimum. For such an orbit the relative distance vector and the relative momentum are perpendicular to each other ($\vartheta = 90^\circ$) and the kinetic energy is given by angular motion only. The quantum mechanical picture is quite different. To lower the total kinetic energy the system delocalizes and spreads over all phase space as reflected in the Wigner function. This is the case even for a soft interaction that shows a low-momentum peak that extends over a large range of distances. In the presence of short-range and tensor correlations the system also extends into the high-momentum region where one can find pairs with high virtuality.


\appendix

\section{Schematic two-Gaussian model}
\label{app:schematic}

To illustrate the properties of the Wigner function and how they are related to the densities in coordinate and momentum space it is helpful to discuss a schematic model with a wave function given as the superposition of a long-ranged and a short-ranged Gaussian. In this simple picture it is easy to disentangle the contributions from short- and long-range (low- and high-momentum) parts of the wave function and it becomes obvious that the interference between these components plays a crucial role.

In the schematic model the total wave function
\begin{equation}
  \psi(\vec{r}) = \alpha_1 \psi_1(\vec{r}) + \alpha_2 \psi_2(\vec{r})
\end{equation}
is given by the superposition of two components that are normalized Gaussians
\begin{equation}
  \psi_i(\vec{r}) = \frac{1}{(\pi a_i)^{3/4}} \exp \left\{ - \frac{\vec{r}^2}{2 a_i} \right\} \: .
\end{equation}
The width parameters $a_1 = 4.0\:\fm^2$ and $a_2 = 0.25\:\fm^2$ are chosen to approximately reproduce the scales of the deuteron wave function. The parameters $\alpha_1 = 1.0315$ and $\alpha_2 = -0.1164$ are determined such that the wave function $\psi(\vec{r})$ is normalized and suppressed at the origin by 90\% compared to that of $\psi_1(\vec{r})$.

In momentum space the wave function is then given by
\begin{equation}
  \tilde{\psi}(\vec{p}) = \alpha_1 \tilde{\psi}_1(\vec{p}) + \alpha_2 \tilde{\psi}_2(\vec{p})
\end{equation}
with the Fourier transformed Gaussians
\begin{equation}
  \tilde{\psi}_i(\vec{p}) = \left(\frac{a_i}{\pi}\right)^{3/4} \exp \left\{ - a_i \frac{\vec{p}^2}{2} \right\} \: .
\end{equation}
At small momenta this wave function is dominated by the long-ranged component $\tilde{\psi}_1(\vec{p})$, 
at high momenta by the short-ranged component $\tilde{\psi}_2(\vec{p})$. As the coefficients $\alpha_1$ and $\alpha_2$ have opposite signs the momentum space wave function will have a node.

\begin{figure}
	\centering
	\includegraphics[width=0.35\textwidth]{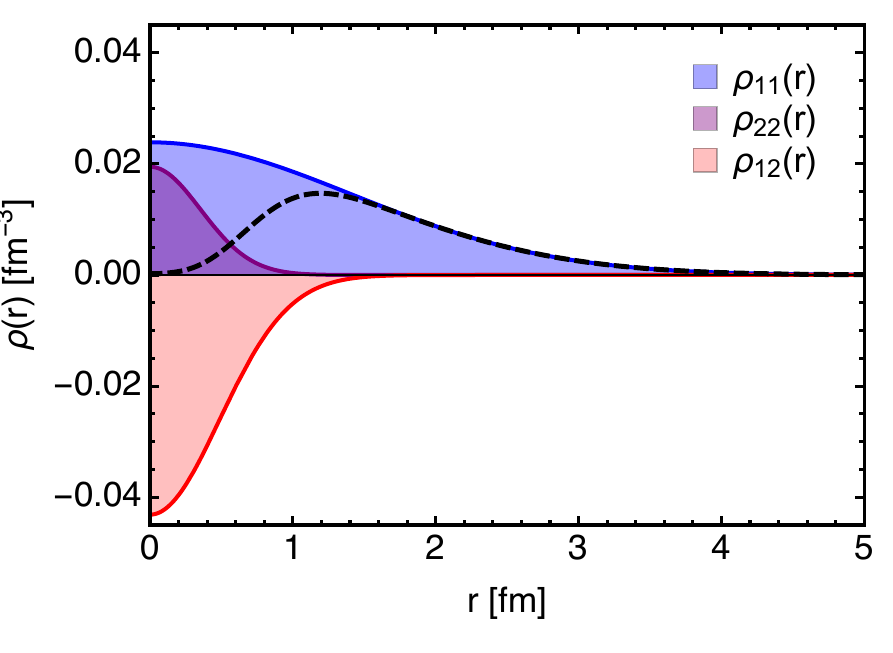}
	\caption{(Color online) Coordinate space density for the schematic two Gaussian wave function. The total density is given by the dashed line. The sum of the diagonal contributions $\rho_{11}(\vec{r})$ and $\rho_{22}(\vec{r})$ are canceled at short distances by the interference term $\rho_{12}(\vec{r})$.}
  \label{fig:coordinate-twogaussians}
\end{figure}

\begin{figure}
	\centering
	\includegraphics[width=0.35\textwidth]{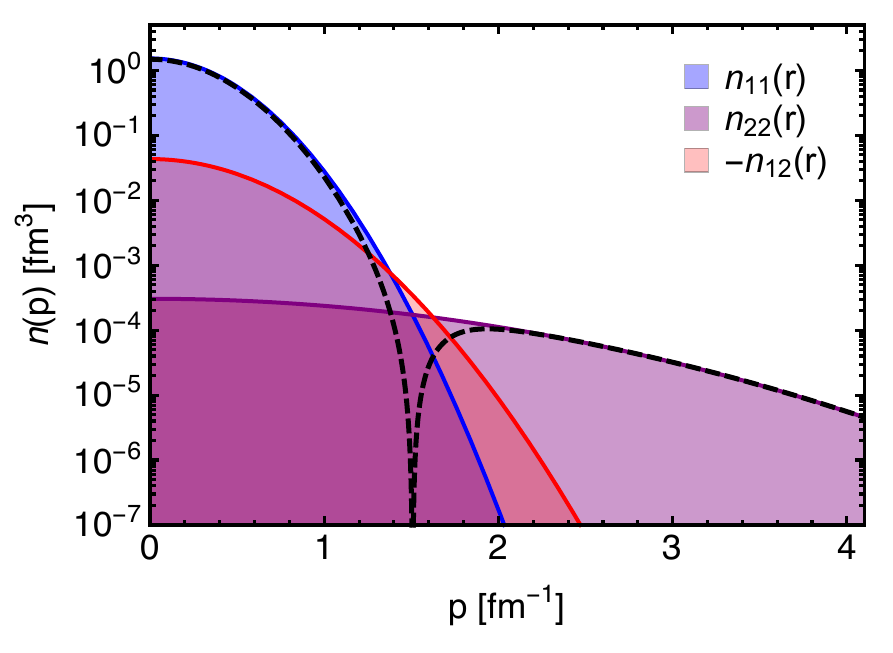}
	\caption{(Color online) Momentum space density for the schematic two Gaussian wave function. The total momentum distribution is given by the dashed line. The low- and high-momentum regions are dominated by the long- and short-ranged Gaussians respectively. The negative contribution of the interference term leads to the dip in the momentum distribution at $p \approx 1.5\:\fm^{-1}$.}
	\label{fig:momentum-twogaussians}
\end{figure}

\begin{figure}
	\centering
	\includegraphics[width=0.33\textwidth]{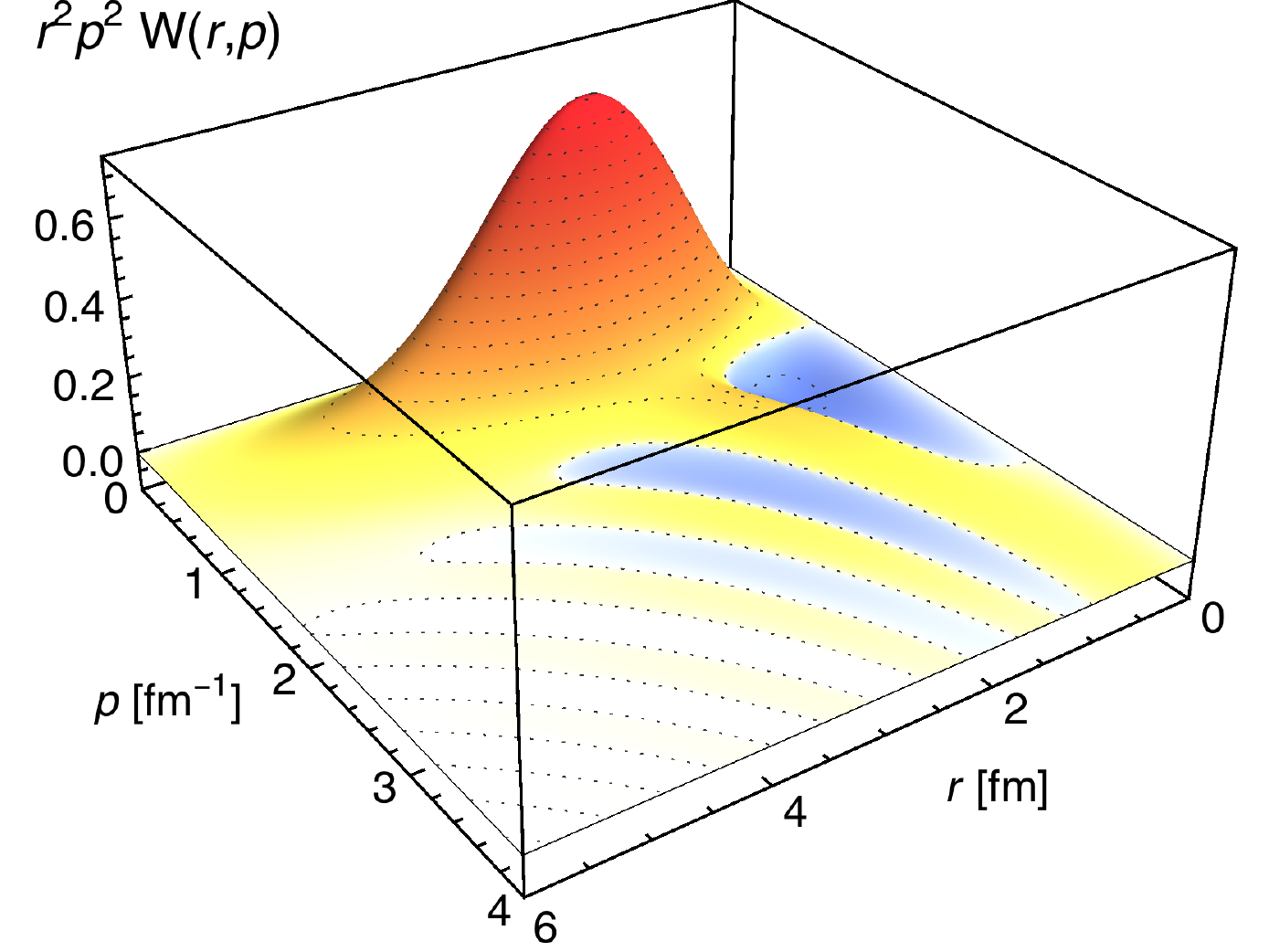}\\
	\includegraphics[width=0.33\textwidth]{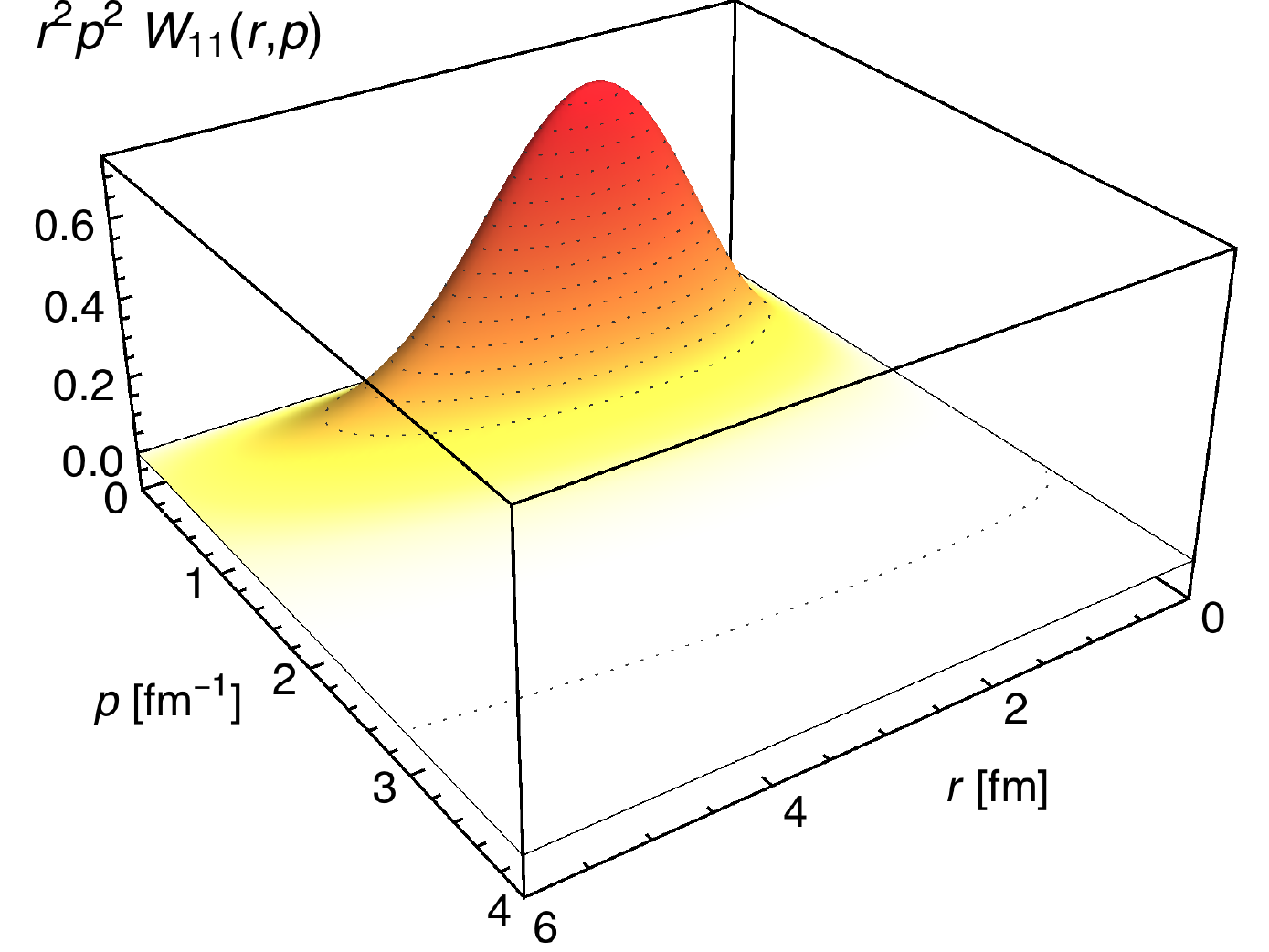}\\
	\includegraphics[width=0.33\textwidth]{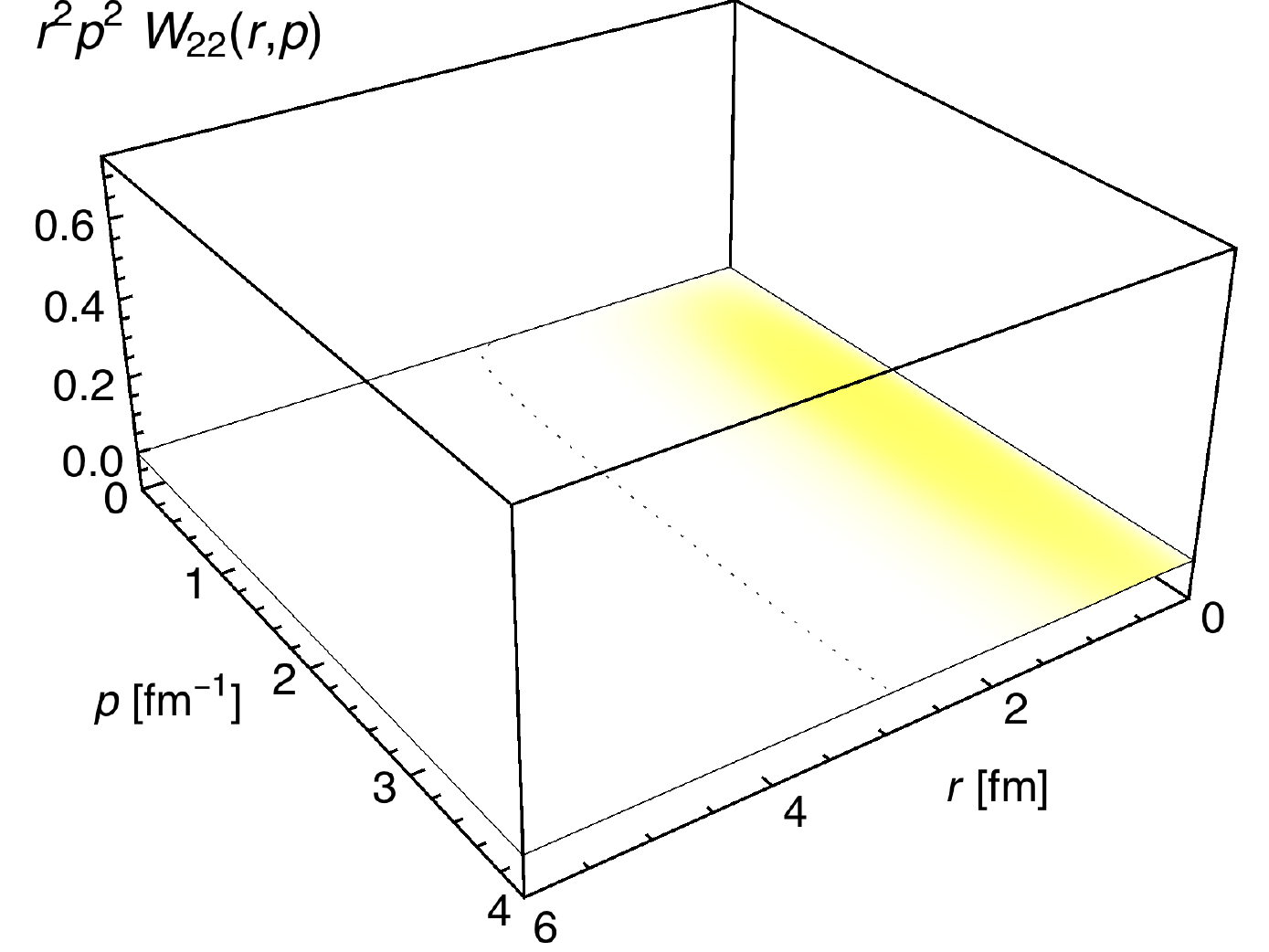}\\
	\includegraphics[width=0.33\textwidth]{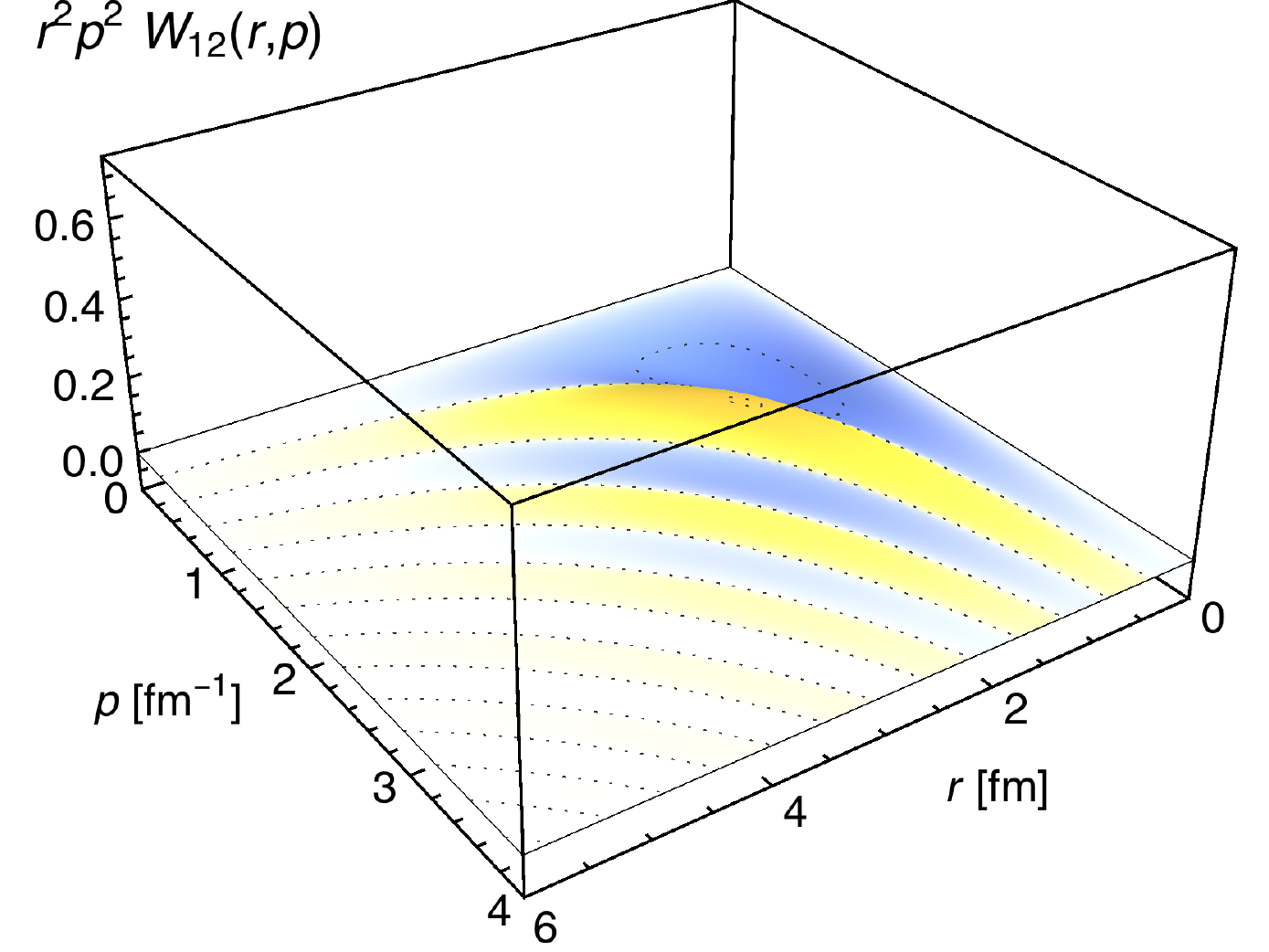}
	\caption{(Color online) (Top) Wigner function $W(r,p)$ scaled with $r^2 p^2$ for the schematic two-Gaussian wave function, followed by the diagonal contribution of the long-ranged Gaussian $W_{11}(r,p)$, the diagonal contribution of the short-ranged Gaussian $W_{22}(r,p)$, and the interference term $W_{12}(r,p)$.}
	\label{fig:wigner-twogaussians}
\end{figure}

These properties of the wave function are of course reflected in the coordinate space density shown in Fig.~\ref{fig:coordinate-twogaussians}. The total density is given by
\begin{equation}
  \rho(\vec{r}) = \rho_{11}(\vec{r}) + \rho_{22}(\vec{r}) + \rho_{12}(\vec{r})
\end{equation}
with the positive diagonal contributions of the long- and short-ranged Gaussians
\begin{equation}
  \rho_{11}(\vec{r}) = \left| \alpha_1 \psi_1(\vec{r}) \right|^2, \quad
  \rho_{22}(\vec{r}) = \left| \alpha_2 \psi_2(\vec{r}) \right|^2
\end{equation}
and the negative contribution from the interference term
\begin{equation}
  \rho_{12}(\vec{r}) = \conj{\alpha_1} \alpha_2 \: \conj{\psi_1(\vec{r})} \psi_2(\vec{r}) + c.c. \: .
\end{equation}
Obviously the interference term is not a small correction but plays an essential role at distances up to about 1~fm. The correlation hole is created by the cancelation of the direct and the interference contributions.

The momentum distributions are shown in Fig.~\ref{fig:momentum-twogaussians}. As in coordinate space we decompose the total momentum space density
\begin{equation}
  n(\vec{p}) = n_{11}(\vec{p}) + n_{22}(\vec{p}) + n_{12}(\vec{p})
\end{equation}
in direct and interference contributions. In momentum space the interference term plays an important role only at intermediate momenta. As the momentum space wave function has a node at $p \approx 1.5\:\fm^{-1}$ also the momentum space density has to vanish there. In terms of the densities this is achieved by the cancelation of direct and interference contributions.

The Wigner function for the schematic model can be decomposed in a similar fashion into diagonal and off-diagonal contributions
\begin{equation}
  W(\vec{r},\vec{p}) = W_{11}(\vec{r},\vec{p}) + W_{22}(\vec{r},\vec{p}) + W_{12}(\vec{r},\vec{p}) \: .
\end{equation}
In Fig.~\ref{fig:wigner-twogaussians} we show the total Wigner function $W(r,p)$ and the individual diagonal and off-diagonal contributions. The total Wigner function of the schematic model is similar to that of the deuteron. We find a large low-momentum peak and an oscillating behavior at larger momenta with the `shoulder' at small distances extending to larger momenta. Because of the Gaussian wave functions the diagonal contributions $W_{11}(r,p)$ and $W_{22}(r,p)$ are purely positive and show no oscillations. The Wigner function of the long-ranged Gaussian $W_{11}(r,p)$ essentially gives the full low-momentum part of the full Wigner function. The contribution of the short-ranged Gaussian $W_{22}(r,p)$ is restricted to the short-distance region and its absolute contribution is very small, much smaller than the contribution from the interference term  $W_{12}(r,p)$. The interference term also shows the characteristic oscillatory behavior that can be traced back to the oscillating term Eq.~\ref{eq:wigner-gaussian-interference} in Eq.~\ref{eq:wigner-gaussian}. It explains the origin of the oscillations as interferences between short- and long-range components (different widths $a_m$ and $a_n$) and also their dependence on $\cos \vartheta$. All these basic properties can also be seen in the realistic deuteron Wigner function.

\begin{widetext}

\section{Evaluation of the Wigner function}
\label{app:calculation}

\subsection{Wigner function for $S$-wave functions}

The evaluation of the Wigner function in Eq.~\eqref{eq:wigner} becomes much easier by our choice of Gaussian basis wave functions given in Eq.~\eqref{eq:gaussian}. In the case of pure $S$-wave functions the integration of the spatial part of the Wigner is a simple Gaussian integration that can be performed analytically for the basis functions with width parameters $a_m$ and $a_n$
\begin{equation}
	\begin{split}
	W^{0}_{mn}(\vec{r},\vec{p}) & = \int d^3s \: \frac{1}{4\pi} \exp \left\{ - \frac{(\vec{r}+\tfrac{1}{2}\vec{s})^2}{2 a_m} - \frac{(\vec{r}-\tfrac{1}{2}\vec{s})^2}{2 a_n} - i \vec{p}\cdot\vec{s} \right\} \\
	&= \frac{1}{4\pi} \int d^3s \: \exp \left\{ - \alpha_{mn} \left(\vec{s} - \vec{\beta}_{mn}(\vec{r},\vec{p}) \right)^2 \right\} G_{mn}(\vec{r},\vec{p}) \\
	&= I^0(\alpha_{mn}) \: G_{mn}(\vec{r},\vec{p})
	\end{split}
	\label{eq:wigner-gaussian}
\end{equation}
with
\begin{equation}
	\alpha_{mn} = \frac{1}{8} \left( \frac{a_m+a_n}{a_m a_n} \right) \: , \qquad
  \vec{\beta}_{mn}(\vec{r},\vec{p}) = 2 \frac{a_m-a_n}{a_m+a_n} \vec{r} - 4 i \frac{a_m a_n}{a_m+a_n} \vec{p} \: ,
\end{equation}
\begin{equation}
  G_{mn}(\vec{r},\vec{p}) = \exp \left\{ -\frac{2}{a_m+a_n} \vec{r}^2 - 2i \frac{a_m-a_n}{a_m+a_n} \vec{r}\cdot\vec{p} - 2 \frac{a_m a_n}{a_m+a_n} \vec{p}^2 \right\}
\end{equation}
and
\begin{equation}
	I^0(\alpha) = \int_0^\infty ds \: s^2 \exp \left\{ - \alpha s^2 \right\} = \frac{\sqrt{\pi}}{4 \alpha^{3/2}}
\end{equation}
The oscillating behavior of the Wigner function is caused by the complex term
\begin{equation}
	\exp \left\{  - 2i \frac{a_m-a_n}{a_m+a_n} \vec{r}\cdot\vec{p} \right\} 
	\label{eq:wigner-gaussian-interference}
\end{equation}
in $G_{mn}(\vec{r},\vec{p})$. For large values of $\vec{r}\cdot\vec{p}$ this oscillating part of the Wigner function will be dominated by terms with a combination of a small $a_m$ and a large $a_n$ (or vice versa). If we assume that one width parameter is much larger than the other the oscillating term becomes
\begin{equation}
	\exp \left\{  - 2i \frac{a_m-a_n}{a_m+a_n} \vec{r}\cdot\vec{p} \right\} \approx 
	\exp \left\{  \pm 2 i \vec{r}\cdot\vec{p} \right\} = \exp \left\{  \pm 2 i r p \cos \vartheta \right\} \: .
\end{equation}
This explains the observed patterns in Fig.~\ref{fig:wignerscaledrptheta-av8p}. When integrating over the angles
we obtain the more rapidly falling off oscillating behavior given by
\begin{equation}
	\int_{-1}^{+1} d(\cos \vartheta) \exp \left\{  \pm 2 i r p \cos \vartheta \right\} = \frac{\sin (2\, r p)}{r p}
\end{equation}
that explains the pattern seen in Fig.~\ref{fig:wignerrpscaled}. 

In the full Wigner function the imaginary parts will cancel and only the oscillating real parts survive. For the textbook case of a single Gaussian wave function there is only one width parameter $a$ and the Wigner function becomes particularly simple as it is the product of Gaussians in coordinate and momentum space
\begin{equation}
  W^{0}(\vec{r},\vec{p})
  = \frac{1}{\pi^3} \exp \left\{ - \vec{r}^2/a \right\} \exp \left\{  - a \vec{p}^2 \right\} \: .
\end{equation}
In this special case the Wigner function is positive everywhere and it has been proven that a necessary and sufficient condition for the Wigner function to be nonnegative is to be of Gaussian form \cite{soto83}.

\subsection{Wigner function in the general case}

In the general case one has to evaluate for the spatial part of the Wigner function integrals of the form
\begin{equation}
	W^{(L_1L_2)LM}_{mn}(\vec{r},\vec{p}) = \int d^3s \: \left\{ Y_{L_1}(\vec{r}+\tfrac{1}{2}\vec{s}) \otimes Y_{L_2}(\vec{r}-\tfrac{1}{2}\vec{s}) \right\}_{LM} \exp \left\{ - \frac{(\vec{x}+\tfrac{1}{2}\vec{s})^2}{2 a_m} - \frac{(\vec{x}-\tfrac{1}{2}\vec{s})^2}{2 a_n} - i \vec{p}\cdot\vec{s} \right\} \: ,
\end{equation}
where we have already recoupled the orbital angular momenta $L_1$ and $L_2$ of the bra and ket states. We also make use of solid spherical harmonics
\begin{equation}
	Y_{LM}(\vec{r}) = r^L Y_{LM}(\unitvec{r}) \: ,
\end{equation}
to obtain
\begin{equation}
	\begin{split}
		W^{(L_1L_2)LM}_{mn}(\vec{r},\vec{p}) 
		&= \int d^3s \: \left\{ Y_{L_1}(\vec{r}+\tfrac{1}{2}\vec{s}) \otimes Y_{L_2}(\vec{r}-\tfrac{1}{2}\vec{s}) \right\}_{LM} \exp \left\{ - \alpha_{mn} (\vec{s} - \vec{\beta}_{mn}(\vec{r},\vec{p}))^2 \right\} G_{mn}(\vec{r},\vec{p}) \\
	& = \int d^3s' \: \left\{ Y_{L_1}(\tfrac{1}{2}\vec{s}' + \vec{r} + \tfrac{1}{2} \vec{\beta}_{mn}(\vec{r},\vec{p})) \otimes Y_{L_2}(-\tfrac{1}{2}\vec{s}' + \vec{r} - \tfrac{1}{2} \vec{\beta}_{mn}(\vec{r},\vec{p})) \right\}_{LM} \exp \left\{ - \alpha_{mn} (\vec{s}')^2 \right\} G_{mn}(\vec{r},\vec{p}) \\
	&= \int d^3s' \: \left\{ Y_{L_1}(\tfrac{1}{2}\vec{s}' + \epsilon^r_1 \vec{r} + \epsilon^p_1 \vec{p}) \otimes Y_{L_2}(-\tfrac{1}{2}\vec{s}' + \epsilon^r_2 \vec{r} + \epsilon^p_2 \vec{p}) \right\}_{LM} \exp \left\{ - \alpha_{mn} (s')^2 \right\} G_{mn}(\vec{r},\vec{p})
	\end{split}
\end{equation}
with
\begin{equation}
  \epsilon^r_1 = 2\frac{a_m}{a_m+a_n}, \qquad 
  \epsilon^r_2 = 2\frac{a_n}{a_m+a_n}, \qquad
  \epsilon^p_1 = -2 i \frac{a_m a_n}{a_m+a_n}, \qquad
  \epsilon^p_2 = 2 i \frac{a_m a_n}{a_m+a_n} \: .
\end{equation}
Using the properties of the tripolar harmonics the integral can be solved analytically

\begin{multline}
  W^{(L_1L_2)LM}_{mn}(\vec{r},\vec{p}) = \frac{1}{2\pi^2} (2L_1+1)(2L_2+1) \sqrt{(2L_1)!(2L_2)!} 
  \sum_{\lambda_1=0,\Lambda_1=L_1-\lambda_1,\Lambda_2=L_2-\lambda_1}^{\min\{L_1,L_2\}} 
  (-1)^{\lambda_1} I^{\lambda_1}(\alpha_{mn})
  \sqrt{(2\Lambda_1+1)(2\Lambda_2+1)} \\
  \times \sum_{L_{r_1}=0,L_{p_1}=\Lambda_1-L_{r_1}}^{\Lambda_1}
  \sum_{L_{r_2}=0,L_{p_2}=\Lambda_2-L_{r_2}}^{\Lambda_2}
  \frac{(\epsilon^r_1 r)^{L_{r_1}} (\epsilon^r_2 r)^{L_{r_2}} (\epsilon^p_1 p)^{L_{p_1}} (\epsilon^p_2 p)^{L_{p_2}}}{\sqrt{(2L_{r_1})!(2L_{r_2})!(2L_{p_1})!(2L_{p_2})!}} \\
  \times \sum_{L_r,L_p} (-1)^{L_1+\Lambda_2+L} \cg{L_{r_1}}{0}{L_{r_2}}{0}{L_r}{0} \cg{L_{p_1}}{0}{L_{p_2}}{0}{L_p}{0}
  \sixj{L_2}{L_1}{L}{\Lambda_1}{\Lambda_2}{\lambda_1} \ninej{L_{r_1}}{L_{r_2}}{L_r}{L_{p_1}}{L_{p_2}}{L_p}{\Lambda_1}{\Lambda_2}{L} \left\{ Y_{L_r}(\unitvec{r}) \otimes Y_{L_p}(\unitvec{p}) \right\}_{LM} G_{mn}(\vec{r},\vec{p})
\end{multline}
with
\begin{equation}
  I^{\lambda}(\alpha) = \int_{0}^{\infty} ds \: s^2 \left(\frac{s}{2}\right)^{2\lambda} \exp\left\{-\alpha s^2\right\} = 
  \left(\frac{1}{2}\right)^{2\lambda+1} \frac{(2\lambda+1)!!}{2^{\lambda+1}\alpha^{\lambda+1}} \sqrt{\frac{\pi}{\alpha}}
\end{equation}

\subsection{Tripolar spherical harmonics}
\label{app:tripolarharmonics}

We make use of the following properties of tripolar spherical harmonics as given in \cite{varshalovich}. With $\vec{r} = \vec{r}_1 + \vec{r}_2 + \vec{r}_3$ the solid harmonic $Y_{LM}(\vec{r})$ can be expanded using tripolar spherical harmonics:
\begin{equation}
  Y_{LM}(\vec{r}) = r^L Y_{LM}(\unitvec{r}) = 4\pi \sqrt{(2L+1)!} \sum_{l_1+l_2+l_3=L} 
  \frac{r_1^{l_1} r_2^{l_2} r_3^{l_3}}{\sqrt{(2l_1+1)! (2l_2+1)! (2l_3+1)!}} 
  \left\{ Y_{l_1}(\unitvec{r}_1) \left\{ Y_{l_2}(\unitvec{r}_2) Y_{l_3}(\unitvec{r}_3) \right\}_{L-l_1} \right\}_{LM} \: .
\end{equation}
The recoupled product of tripolar spherical harmonics with the same arguments is given by
\begin{multline}
	\left\{ \left\{ Y_{l_1'}(\unitvec{r}_1) \left\{ Y_{l_2'}(\unitvec{r}_2) Y_{l_3'}(\unitvec{r}_3) \right\}_{\lambda'} \right\}_{L'} \left\{ Y_{l_1''}(\unitvec{r}_1) \left\{ Y_{l_2''}(\unitvec{r}_2) Y_{l_3''}(\unitvec{r}_3) \right\}_{\lambda''} \right\}_{L''} \right\}_{LM} = \\
	\sum_{l_1 l_2 l_3 \lambda} B^{l_1 l_2 l_3 \lambda L}_{l_1' l_2' l_3' \lambda' L' l_1'' l_2'' l_3'' \lambda'' L''} 
	\left\{ Y_{l_1}(\unitvec{r}_1) \left\{ Y_{l_2}(\unitvec{r}_2) Y_{l_3}(\unitvec{r}_3) \right\}_{\lambda} \right\}_{LM}
\end{multline}
with
\begin{multline}
	B^{l_1 l_2 l_3 \lambda L}_{l_1' l_2' l_3' \lambda' L' l_1'' l_2'' l_3'' \lambda'' L''} =
		\frac{1}{(4\pi)^3} \sqrt{(2l_1'+1)(2l_1''+1)(2l_2'+1)(2l_2''+1)(2l_3'+1)(2l_3''+1)(2L'+1)(2L''+1)} \\
		\times \sqrt{(2\lambda'+1)(2\lambda''+1)(2\lambda+1)} \cg{l_1'}{0}{l_1''}{0}{l_1}{0} \cg{l_2'}{0}{l_2''}{0}{l_2}{0} \cg{l_3'}{0}{l_3''}{0}{l_3}{0}
		\ninej{l_1'}{l_1''}{l_1}{\lambda'}{\lambda''}{\lambda}{L'}{L''}{L} 
		\ninej{l_2'}{l_2''}{l_2}{l_3'}{l_3''}{l_3}{\lambda'}{\lambda''}{\lambda}
\end{multline}
In the special case $l_1=0$ (corresponding to the $\vec{s}'$ variable) this simplifies with $l_1' = l_1''$ and $\lambda = L$. We get
\begin{multline}
	B^{0 l_2 l_3 L L}_{l_1' l_2' l_3' \lambda' L' l_1' l_2'' l_3'' \lambda'' L''} =
		(-1)^{L'+\lambda''+L} \frac{1}{(4\pi)^3} \sqrt{(2l_2'+1)(2l_2''+1)(2l_3'+1)(2l_3''+1)(2L'+1)(2L''+1)} \\
		\times \sqrt{(2\lambda'+1)(2\lambda''+1)} \cg{l_2'}{0}{l_2''}{0}{l_2}{0} \cg{l_3'}{0}{l_3''}{0}{l_3}{0}
		\sixj{L''}{L'}{L}{\lambda'}{\lambda''}{l_1'} 
		\ninej{l_2'}{l_2''}{l_2}{l_3'}{l_3''}{l_3}{\lambda'}{\lambda''}{L} \: .
\end{multline}

\end{widetext}

\bibliography{wigner}

\end{document}